\RequirePackage{amsmath, amssymb, bm}
\documentclass[10pt]{iopart}

\usepackage{graphicx}
\usepackage[
hidelinks,
colorlinks=true,
urlcolor=blue,
citecolor=blue,
linkcolor=blue,
]{hyperref}
\usepackage{doi}
\usepackage{natbib}
\usepackage{lineno}
\usepackage{my}

\usepackage{orcidlink}

\usepackage[norefs, nocites, ignoreunlbld]{refcheck}

\begin{document}

\title[Hasegawa et al.]{Continuous data assimilation of large eddy simulation by lattice Boltzmann method and local ensemble transform Kalman filter (LBM-LETKF)}

\author{
Yuta Hasegawa$^{1,*}$\orcidlink{0000-0002-4072-6311},
Naoyuki Onodera$^1$\orcidlink{0000-0001-7392-2899},
Yuuichi Asahi$^1$\orcidlink{0000-0002-9997-1274},
Takuya Ina$^1$\orcidlink{0000-0002-3989-5011},
Toshiyuki Imamura$^2$\orcidlink{0000-0003-1601-9710},
and
Yasuhiro Idomura$^1$\orcidlink{0000-0002-2829-0498}
}
\address{
    $^1$
    Center for Computational Science and e-Systems,
    Japan Atomic Energy Agency,
    178-4-4-4F Wakashiba, Kashiwa-shi, Chiba 277-0871, Japan\\
    $^2$
    RIKEN Center for Computational Science,
    7-1-26 Minatojima-minami-machi, Chuo-ku, Kobe, Hyogo 650-0047, Japan
}
\ead{hasegawa.yuta@jaea.go.jp}

\begin{abstract}
We investigate the applicability of the data assimilation (DA) to large eddy simulations (LESs) based on the lattice Boltzmann method (LBM).
We carry out the observing system simulation experiment of a two-dimensional (2D) forced isotropic turbulence, and examine the DA accuracy of the nudging and the local ensemble transform Kalman filter (LETKF) with spatially sparse and noisy observation data of flow fields.
The advantage of the LETKF is that it does not require computing spatial interpolation and/or an inverse problem between the macroscopic variables (the density and the pressure) and the velocity distribution function of the LBM, while the nudging introduces additional models for them.
The numerical experiments with $256\times256$ grids and 10\% observation noise in the velocity showed that the root mean square error of the velocity in the LETKF with $8\times 8$ observation points ($\sim 0.1\%$ of the total grids) and 64 ensemble members becomes smaller than the observation noise, while the nudging requires an order of magnitude larger number of observation points to achieve the same accuracy. 
Another advantage of the LETKF is that it well keeps the amplitude of the energy spectrum, while only the phase error becomes larger with more sparse observation.
We also see that a lack of observation data in the LETKF produces a spurious energy injection in high wavenumber regimes, leading to numerical instability. 
Such numerical instability is known as the catastrophic filter divergence problem, which can be suppressed by increasing the number of ensemble members.
From these results, it was shown that the LETKF enables robust and accurate DA for the 2D LBM with sparse and noisy observation data.


\noindent\preprint

\end{abstract}

\submitto{\FDR}
\maketitle

\clearpage
\section{Introduction}

\begin{figure}[b]
    \centering
    \includegraphics[width=\linewidth]{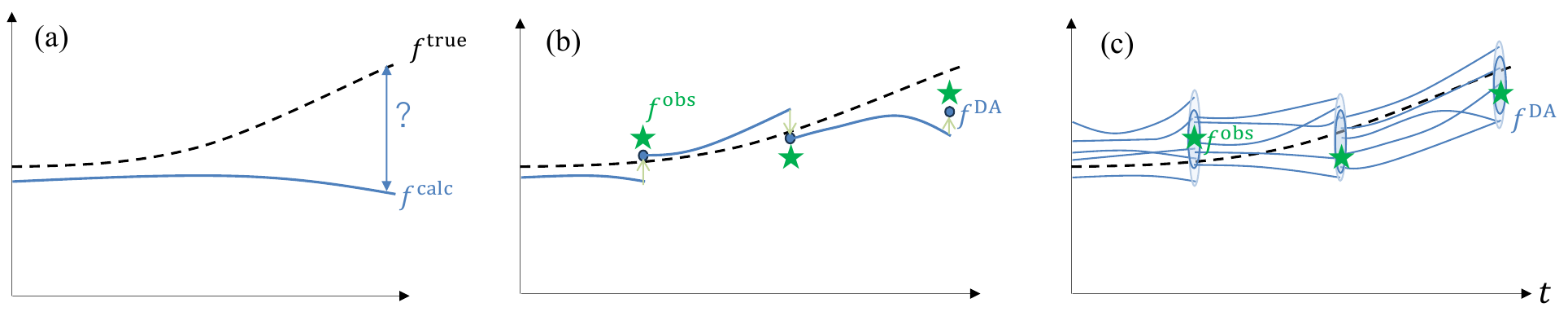}
    \caption{
    Conceptual schematics of the continuous DA for a chaotic system with noisy observation.
    (a) Single-shot simulation without DA, which may predict the wrong state due to the chaotic nature of the system.
    $f^\mathrm{true}$ and $f^\mathrm{calc}$ are the quantities of the true state and the numerical simulation, respectively.
    (b) Single-shot simulation with the DA (typically, by the nudging method).
    $f^\mathrm{obs}$ is the observed data from the true state with some noises.
    $f^\mathrm{DA}$ is the predicted state improved by the DA.
    (c) Ensemble simulation with ensemble Kalman filter.
    The filled circles show ensemble variation, where the thin and thick filling colors respectively indicate the range of ensemble data before and after the DA.
    }
    \label{fig:enkf_concept}
\end{figure}

The data assimilation (DA) refers to the methodology which interpolates the simulation and the experiment to achieve more consistent state estimation.
The \textit{continuous} DA is one of the major DA fields,
in which observation and DA are repeated in time (cf. \figref{fig:enkf_concept} (a), (b)).
The continuous DA is used to estimate and/or predict the time-dependent state of chaotic systems, which have been the major interest in the community of the numerical weather prediction (NWP)~\citep{Kalnay2002}.
Recently, the continuous DA, also known as the reconstruction, regeneration, or state estimation of flow fields, has also been applied to the field of computational fluid dynamics (CFD) to predict turbulent flows.
Fundamental properties of DA in turbulent flows have been investigated in direct numerical simulations (DNSs) of homogeneous and isotropic turbulence. 
It was shown that by imposing a reference solution to larger-scale Fourier modes below a certain critical wavenumber, 3D DNSs of isotropic turbulence were fully reproduced or synchronized up to smaller-scale Fourier modes~\citep{Yoshida2005,Lalescu2013,Wang2022}.
The simplest continuous DA approach called the \textit{nudging}~\citep{Hoke1976,Lakshmivarahan2013-hyphen} was applied to 2D isotropic turbulence, and conditions for synchronization, in which the assimilated solution converges to the reference solution in time, were shown for nudging in Fourier space~\citep{Olson2003} and in configuration space~\citep{Azouani2014}, and were examined in 2D DNSs of isotropic turbulence~\citep{Olson2008,Gesho2016}. 
The DA accuracy of the nudging was investigated also in 3D DNSs of isotropic turbulence, and it was shown that to obtain synchronized solutions, $\sim 20\%$ of volume or grid points have to be assimilated in configuration space, while nudging in Fourier space leads to synchronized solutions with an order of magnitude lower number of nudged Fourier modes~\citep{Leoni2020}.
\citet{Li2022a} also studied the nudging DA of 3D isotropic turbulence with the large eddy simulation (LES).

The nudging based on flow field data was also applied to turbulent flows around a square cylinder, where 2D unsteady Reynolds-averaged Navier–Stokes (URANS) simulations were assimilated with the reference flow field data from 3D DNSs \citep{Zauner2022}.
Although many of these works addressed an impact of observation resolution in configuration space or in Fourier space on the DA accuracy, an influence of observation noise was studied only in a few works.
\citet{Wang2022a} investigated the nudging DA for the turbulent channel flow with observation noises;
however, in the latter work, the gain of the nudging was set to 1, i.e., the simulation state was totally replaced by the observation data where they exist.
Theoretically, it is difficult to determine the optimal gain of the nudging against given observation/simulation errors, and thus, the gain of the nudging is determined empirically in general. 

On the other hand, the issue of observation error was addressed also by using Kalman filter approaches. 
\citet{Suzuki2012} proposed a hybrid simulation combining particle tracking velocimetry (PTV) and DNS, in which 2D DNSs of a planar-jet flow were assimilated with PTV data from the experiment using the reduced-order Kalman filter.
\citet{Colburn2011} showed the capability of the ensemble Kalman filter (EnKF)~\citep{Evensen1994a,Evensen2003a} applied to the flow reconstruction in the turbulent channel flow with the measurement of the pressure on the walls.
With increasing computing power, continuous DAs were also applied to engineering problems in recent years.
\citet{Bauweraerts2021} showed a turbulent flow reconstruction in the atmospheric boundary layer by the 4D-Var DA~\citep{LeDimet1986} with the (virtual) lidar measurement.
\citet{Labahn2020} applied the EnKF to 3D LES of a turbulent jet, which was assimilated with flow fields reconstructed based on tomographic particle image velocimetry (TPIV) data from the experiment.
They examined its DA accuracy by changing DA conditions such as the number of ensemble members, the localization parameter, the observation error, the DA frequency, and the resolution of reconstructed flow fields.

Although the latter study~\citep{Labahn2020} demonstrated promising features of the EnKF in engineering problems, the analysis was still {\it posteriori}, and the observation resolution was relatively high (several experimental data points per computational cell).
As the goal of DA is to produce accurate forecasts based on observation data, real-time or even faster CFD simulations are needed.
Although the space and time resolution of the experimental data is dramatically improving with PIV techniques in the laboratory experiment, it may still be limited in the open fields such as atmospheric boundary layer flows. 
Therefore, one needs to examine the applicability of the continuous DA to more sparse observation conditions.
To address these two issues, in this work, we apply the continuous DA to the lattice Boltzmann method (LBM), and address its DA accuracy with noisy and spatially-sparse observation conditions.
Here, the LBM is an essential tool for the large-scale LES, in which over billion degree-of-freedoms (DOFs) can be computed on state-of-the-art supercomputers.
In our previous works~\citep{Onodera2018, Onodera2021}, for instance, the LBM was dramatically accelerated on GPUs, and real-time ensemble LESs for plume dispersion analysis in urban areas were enabled by combining the LBM and the block-structured adaptive mesh refinement (AMR) method.
Regarding the continuous DA, we used the nudging in the latter work.
However, in the nudging, spatial interpolation is needed when the simulation and the observation have different resolutions.
In addition, the nudging requires the conversion of the variables, when the variables of the simulation and the observation are different; namely, the LBM solves the velocity distribution function, while the macroscopic variables such as the velocity and the pressure are observed.
Here, the conversion from the macroscopic variables to the velocity distribution cannot be uniquely determined, and an approximation such as assuming the equilibrium distribution function is needed. 
Such approximations can be naturally avoided by using the EnKF, in which the state vector and the observation vector can be designed by using different variables at different resolutions and dimensions.
However, the applicability of the EnKF to the LBM has not been tested to our best knowledge.
Thus, in this work, we clarify this point by comparing with our previous method, the nudging.

Here, to consider the candidate of the continuous DA approaches applied to the LBM, we briefly review methodologies developed in the NWP community.
The continuous DA can be classified into two major approaches: one is the variational approach and the other is the statistical approach.
Historically, in the former, the 3D or 4D variational method (3D/4D-Var) has been the \textit{de facto} standard DA approach.
However, the 3D/4D-Var has difficulty in determining the background error (i.e. the error and the covariance of the numerical state are prescribed by the model hyperparameters).
The determination of the background error is often based on an empirical rule, which varies depending on the numerical model.
One needs to redesign the background error whenever the model is changed, e.g., adjusting the domain size, the mesh resolution, the time step width, boundary conditions, and/or switching numerical schemes such as space discretization and time integration.
Such frequent redesign of the background error is very expensive and is a serious problem in mesh refinement approaches, in which the mesh resolution changes in time.
In addition, especially in the 4D-Var, one needs to implement the so-called \textit{adjoint} code, which is another critical issue for the adaptive mesh refinement approaches.

On the other hand, another major paradigm, which can avoid the redesign of the background error, is the EnKF.
The EnKF computes the ensemble simulation, in which the statistics of the numerical model are estimated by multiple simulations with slightly different simulation conditions (cf. \figref{fig:enkf_concept} (c)).
The ensemble simulation is regarded as the Monte Carlo sampling, and thus, its statistical error trivially corresponds to the background error.
This is a more straightforward approach than the 3D/4D-Var because an empirical design of the background error is no longer needed. Therefore, the applicability of the EnKF is drastically enhanced from the 3D/4D-Var, and we address the EnKF in this work.

Since the original EnKF~\citep{Evensen1994a} had the problem of too small ensemble variation~\citep{Burgers1998}, the EnKF has been improved in former studies. They are classified mostly into two types:
One is the perturbed observation method (EnKF-PO)~\citep{Burgers1998}, which superimposes the additional ensemble perturbation onto the observation to implicitly maintain the ensemble variation.
The other is the ensemble squared root filter (EnSRF)~\citep{Tippett2003}, which explicitly updates the separation of each ensemble member by solving the squared root of the ensemble covariance matrix.
Although the relative merits of the two approaches are still under debate, in this study, we choose one of the EnSRF approaches, the local ensemble transform Kalman filter (LETKF)~\citep{Hunt2007a}.
The reason for the decision is that 
the LETKF can be computed in an embarrassingly parallel manner at each grid point,
and thus, is suitable for high-performance computing (HPC).
In fact, the LETKF has been applied to extreme-scale DA simulations with large ensemble sizes and/or DOFs on state-of-the-art supercomputers,
e.g., on the CPU-based supercomputers in the former works~\citep{Miyoshi2014,Miyoshi2016a,Yashiro2020},
and on the GPU-based supercomputer in our previous study~\citep{Hasegawa2022}. 

The LETKF and the other EnKFs have been well-studied in the field of CFD, however,
a few works of those were conducted with the LBM.
\citet{Roussel2013} estimated unknown boundary conditions in the 2D lid-driven cavity flow problem by using the EnKF-PO with the velocity measurements.
\citet{Salman2022} recovered the spatial profile of the temperature in the simulated indoor air quality with unknown boundary temperature by using the 3D-Var or the EnKF-PO with the temperature measurements within the room. 
Those studies, however, utilized the DA for the estimation of the unknown boundary condition rather than the flow state estimation.
Thus, in this study, we apply the LETKF to the LBM, and examine its capability in turbulent flow reconstructions.

As the first test case of the LBM-LETKF, we consider a minimal benchmark problem designed to minimize the requirements on physics modelings and computational costs while keeping chaotic behavior.
For example, in the NWP field,
the NCEP GFS model~\citep{Szunyogh2005a} and the SPEEDY model~\citep{Molteni2003} are the minimal 3D global atmospheric models whose DOFs are typically $\mathcal{O}(10^4)$.
We also consider a small system with similar DOFs. 
In this work, we choose a 2D forced isotropic turbulence~\citep{Lilly1969}, in which properties of the nudging have been well understood in former works.
We also apply an LES model to suppress the grid number to $256\times256$ for computing isotropic turbulence at a relatively high Reynolds number.
Under these conditions, the ensemble simulation with up to 64 ensembles can be computed within a few minutes on GPUs,
and thus, we can conduct thousands of trial-and-error experiments with reasonable computing resources.

The remainder of this paper is organized as follows.
The numerical method of the LBM and the LETKF are described in \sref{s:scheme} and \sref{s:da}, respectively.
While the validation of the LETKF is the major objective of this study,
a simpler DA model, the nudging, is also introduced for comparison.
In the LBM, the state and observable variables are different, i.e., the former is the velocity distribution function and the latter is the macroscopic variables of the flow field.
We also describe how to treat such state and observable variables in the LETKF or the nudging.
\sref{s:exp} presents the DA experiment of 2D forced isotropic turbulence with the LBM-LETKF.
We first design and validate the numerical configuration, and then, the DA experiments using the LBM-LETKF and the LBM-nudging are compared.
As the observation data, we assume spatially-sparse and noisy observation data of velocity and pressure field, which may be considered for real measurements of the atmospheric boundary layer environment.
Especially, data sparsity will be studied in detail to investigate the accuracy and robustness of the LETKF and to show the advantage of the LETKF compared to the simple nudging method.
Finally, we conclude this study in \sref{s:conclusion}.

\section{Numerical model}
\label{s:scheme}

We start with the definition of the CFD model without the DA scheme.
Since our future target is to implement the ensemble DA into the LBM-based real-time urban wind simulation code~\citep{Onodera2021}, 
we employ the LBM as the CFD model for this study.
Although our previous study utilized the 3D cumulant model~\citep{Geier2015a},
in this study we use the traditional 2D lattice BGK model~\citep{Qian1992} for simplicity.
In the LBM with D2Q9 (two-dimensional nine-velocities) model,
the discrete velocity $\bm c_\alpha$ is defined as 

\begin{equation}
    \bm c_\alpha = 
    \left\{\begin{array}{ll}
        (0, 0)^\top, & \alpha = 0 \\
        (\pm c, 0)^\top, & \alpha = 1, 2 \\
        (0, \pm c)^\top, & \alpha = 3, 4 \\
        (\pm c, \pm c)^\top, & \alpha = 5, 6, 7, 8
    \end{array}\right.,
\end{equation}

\noindent
where the subscript $\alpha$ is the index of the discrete velocity.
$c$ is the model constant called the lattice speed, which is determined from the given grid spacing $\Delta x$ and time step width $\Delta t$, namely, $c = \Delta x / \Delta t$.
The governing equation is defined as

\begin{equation}
    f_\alpha(\bm \xi+\bm c_\alpha\Delta t, t+\Delta t) - f_\alpha(\bm \xi, t) =
    \Delta t\left[
    - \frac{1}{\tau}\left(
        f_\alpha(\bm \xi,t) - f_\alpha^\mathrm{eq}(\bm \xi,t)
    \right)
    + F_\alpha
    \right],
\end{equation}

\noindent
where $\bm \xi=(x,y)$ and $t$ are the coordinate and the time, respectively.
$f_\alpha$ is the velocity distribution function, whose moments correspond to the macroscopic variables as

\begin{eqnarray}
    \rho &= \sum_\alpha f_\alpha, \\
    \bm u &= \frac{1}{\rho}
        \sum_\alpha f_\alpha \bm c_\alpha,
\end{eqnarray}

\noindent
where $\rho$ is the fluid density, and $\bm u = (u, v)$ is the fluid velocity.
$f^\mathrm{eq}_\alpha$ is the equilibrium distribution function, defined by

\begin{equation}
    f^\mathrm{eq}_\alpha = \rho w_\alpha \left( 
    1 + 3\frac{\bm c_\alpha \cdot \bm u}{c^2}
    + \frac{9}{2}\frac{(\bm c_\alpha \cdot \bm u)^2}{c^4} - \frac{3}{2}\frac{\bm u\cdot\bm u}{c^2}
    \right),
\end{equation}

\noindent
where $w_\alpha$ is the weighting factor, which is given by

\begin{equation}
    w_\alpha = \left\{
        \begin{array}{ll}
            4/9, & \alpha = 0 \\
            1/9, & \alpha = 1,2,3,4 \\
            1/36, & \alpha = 5,6,7,8
        \end{array}\right..
\end{equation}

\noindent
$\tau$ is the relaxation time.
From Chapman-Enskog expansion~\citep{Kruger2017}, given the lattice speed $c$ and the time step width $\Delta t$, the relaxation time is chosen to satisfy

\begin{equation}
    \nu + \nu_\mathrm{SGS} = \frac{c^2}{3}\left(\tau - \frac{\Delta t}{2}\right),
\end{equation}

\noindent
where $\nu$ is the kinetic viscosity and $\nu_\mathrm{SGS}$ is the sub-grid scale (SGS) viscosity.
The SGS viscosity is computed based on the standard Smagorinsky model~\citep{Smagorinsky1963a} because the benchmark problem in this study is the isotropic turbulence and thus the velocity field is statistically homogeneous.
$\nu_\mathrm{SGS}$ is then defined as

\begin{eqnarray}
    \nu_\mathrm{SGS} &= (C_s\bar{\Delta})^2|S|,\\
    |S| &= \sqrt{2 S_{ij} S_{ji}}, \\
    S_{ij} &= \frac{1}{2}\left(
        \frac{\partial u_i}{\partial \xi_j} + \frac{\partial u_j}{\partial \xi_i}
    \right).
\end{eqnarray}

\noindent
Here, $C_s=0.2$ is the Smagorinsky constant, and
$\bar{\Delta}$ is the model constant called the filter length, which is set to $\bar{\Delta} = \Delta x$ in this study.
$S_{ij}$ is the strain-rate tensor with the subscripts $i,j$ being the indices of spatial axes.
$F_\alpha$ is the forcing term, which is defined with the given acceleration $\bm F$ by

\begin{equation}
    F_\alpha = 3w_\alpha \rho \frac{\bm F\cdot \bm c_\alpha}{c^2}.
\end{equation}

\section{Continuous data assimilation models}
\label{s:da}

The continuous data assimilation (DA) corrects model variables in the numerical simulation based on the observation data iteratively in time.
We introduce several vectors to treat the numerical and observation data in the context of the DA:
let $\bm x(\bm \xi)$ and $\bm y(\bm \eta)$ be respectively the vector of numerical state (state vector) at the grid space $\bm \xi$, and the vector of observable variables (observation vector) at the grid space $\bm\eta$.
The number of grid points in the observation space $\bm\eta$ is usually much smaller than that in the simulation state $\bm\xi$. In this study, the state vector and observation vector are respectively chosen as the distribution function of the LBM and the macroscopic variables, namely,
\begin{equation}
    \label{eq:x}
    \bm x(\bm\xi) = [f_0(\bm\xi), f_1(\bm\xi), \dots, f_8(\bm\xi)]^\top,
\end{equation}
\begin{equation}
    \label{eq:y}
    \bm y(\bm\eta) = [\rho(\bm\eta), u(\bm\eta), v(\bm\eta)]^\top.
\end{equation}

Let $\bm x^\mathrm{f}$, $\bm x^\mathrm{a}$, and $\bm y^\mathrm{o}$ be respectively the \textit{forecast} (prior) and the \textit{analysis} (posterior) state vectors in the numerical model, and the \textit{observation} vector (measurement data) from the true state at the time $t$.
Hereafter, the superscripts `f', `a', and `o' are used also for other variables in the same convention.
The workflow of the continuous DA, which repeats the cycle of the time integration of the numerical model and the DA, consists of the following procedures:

\begin{enumerate}
    \item (Initial condition) Set the initial analysis state vector, $\bm x^\mathrm{a}(\bm\xi)|_{t=0}$.
    Set $t=0$.

    \item (Forecast) Compute the time integration of the numerical model and get the forecast state vector in the next time step,
    $\bm x^\mathrm{f}(\bm\xi)|_{t=t+s\Delta t}$.
    Here, the number of time steps per the observation period is chosen as $s=200$ in this study.
    The reason for this choice is based on the maximum Lyapunov exponent of the turbulence, which will be discussed in \sref{ss:lyap}.
    \item (Observation) Measure the ground truth state and get the observation vector, 
    $\bm y^\mathrm{o}(\bm\eta)|_{t=t+s\Delta t}$.

    \item (Analysis) Assimilate the forecast state vector with the observation vector and get the new analysis state vector based on the linear estimation theory~\citep{Talagrand1997,Desroziers2006}:

    \begin{equation}
        \bm x^\mathrm{a} = \bm x^\mathrm{f} - \bm K(\bm y^\mathrm{o} - \mathcal{H}\bm x^\mathrm{f}),
    \end{equation}

    \noindent
    $\bm K$ is the gain matrix, which varies depending on the DA method.
    $\mathcal{H}$ is the observation operator defined as $\bm y = \mathcal{H}\bm x$.

    \item Update the time step, $t \leftarrow t+s\Delta t$, and repeat the procedures (i--iv).
\end{enumerate}

\noindent
In practice, once the vectors $\bm x$ and $\bm y$ are designed, the observation operator $\mathcal{H}$ is designed to satisfy $\bm y = \mathcal{H}\bm x$, and the gain $\bm K$ is finally given.
The design of $\bm x, \bm y$, and $\bm K$ in each DA method is described in the following subsections.

\subsection{Nudging}

Firstly, we introduce a traditional DA approach, which is called nudging~\citep{Hoke1976, Lakshmivarahan2013-hyphen}, as a reference in this work.
The nudging gives the simplest linear interpolation between the state and the observation,
i.e.,

\begin{equation}
    \label{eq:nud}
    \bm K =\gamma\mathcal{H}^{-1},
\end{equation}

\noindent
where $0\leq\gamma\leq1$ is an empirical scalar constant.
In this study, we choose optimal $\gamma$ in each case by the preliminary numerical experiment.
Since the nudging method is an elementwise DA procedure, 
the state and the observation can be treated as scalars rather than vectors.
$\mathcal{H}^{-1}$ is the inversion of the observation operator, namely, $\bm x = \mathcal{H}^{-1}\bm y$. To compute $\mathcal{H}^{-1}\bm y$,
one needs \textit{a priori} data processing which transforms from the macroscopic variables to the velocity distribution function.
However, such transformation is not unique.
In this study, we assume that the observation is given as the equilibrium distribution, and the non-equilibrium components are ignored.
In addition, one also needs spatial interpolation for the observation data if the resolution of the observation is coarser than that of the simulation.
In this study, we apply simple linear interpolation to the sparse observation data.
Although these approximations may ruin the DA accuracy, the nudging has merits in its computing cost and ease of use.
Finally, the nudging for the LBM in this study is formed as
\begin{equation}
    f^\mathrm{a}_\alpha(\bm\xi) = f^\mathrm{f}_\alpha(\bm\xi) + \gamma \big[f^\mathrm{eq}_\alpha|_{\rho=\hat{\rho}^\mathrm{o}(\bm\xi), u=\hat{u}^\mathrm{o}(\bm\xi), v=\hat{v}^\mathrm{o}(\bm\xi)} - f^\mathrm{f}_\alpha(\bm\xi)\big],
\end{equation}
where $\hat{\rho}^\mathrm{o}(\bm\xi)$, $\hat{u}^\mathrm{o}(\bm\xi)$, and $\hat{v}^\mathrm{o}(\bm\xi)$ are spatially interpolated variables of the observation $\rho^\mathrm{o}(\bm\eta)$, $u^\mathrm{o}(\bm\eta)$, and $v^\mathrm{o}(\bm\eta)$, respectively.

\subsection{Local ensemble transform Kalman filter (LETKF)\label{ss:letkf}}

The local ensemble transform Kalman filter (LETKF)~\citep{Hunt2007a} is one of the advanced models of the continuous DA, which is a derived form of the ensemble Kalman filter (EnKF)~\citep{Evensen2003a}.
In the LETKF, the DA is defined along with the ensemble simulation.
One of its advantages is that the gain $\bm K$ is computed from the statistics of the ensemble simulation, and the adjoint of observation operator $\mathcal{H}^\dagger$ is thus not required.
It is noted that the observation operator $\mathcal{H}$ and the gain $\bm K$ are not explicitly given for simplicity.
To see their step-by-step deviations, readers can refer to, e.g., the book by \citet{Li2012}.

\subsubsection{Basic model}

Let $M$ be the number of ensemble members (i.e. the ensemble size), and $\bm x_m^\mathrm{f}$ be the forecast state vector of the $m$-th ensemble member, where $m=0,1,\dots, M-1$.
Their ensemble mean and perturbation are then introduced as

\begin{eqnarray}
    \bar{\bm x}^\mathrm{f} &= \frac{1}{M}\sum_{m=0}^{M-1} \bm x_m^\mathrm{f},\\
    \delta \bm X^\mathrm{f} &= [
        \bm x_0^\mathrm{f} - \bar{\bm x}^\mathrm{f},
        \bm x_1^\mathrm{f} - \bar{\bm x}^\mathrm{f},
        \dots,
        \bm x_{M-1}^\mathrm{f} - \bar{\bm x}^\mathrm{f}
        ].
\end{eqnarray}

\noindent
Similarly, those mapped to the observation space are defined by

\begin{eqnarray}
    \bar{\bm y}^\mathrm{f} &= \frac{1}{M}\sum_{m=0}^{M-1} \bm y_m^\mathrm{f},\\
    \delta \bm Y^\mathrm{f} &= [
        \bm y_0^\mathrm{f} - \bar{\bm y}^\mathrm{f},
        \bm y_1^\mathrm{f} - \bar{\bm y}^\mathrm{f},
        \dots,
        \bm y_{M-1}^\mathrm{f} - \bar{\bm y}^\mathrm{f}
        ],
\end{eqnarray}

\noindent
where $\bm y^\mathrm{f}_m={\cal H}\bm x^\mathrm{f}_m$.

The analysis state vectors are solved in a latent ensemble space.
Let $\bm w^\mathrm{a}_m\in\mathbb{R}^M$ be the analysis state vector in the ensemble space, which is given by the transform $\bm x^\mathrm{a}=\bm X^\mathrm{f}\bm w^\mathrm{a}$.
Its ensemble mean $\bar{\bm w}^\mathrm{a}$ and perturbation matrix $\delta \bm W^\mathrm{a}$ are then computed by

\begin{eqnarray}
    \label{eq:letkf_barw}
    \bar{\bm w}^\mathrm{a} &= \tilde{\bm P}^\mathrm{a} \, \delta \bm Y^\mathrm{f\top} \bm R^{-1} 
        (\bm y^\mathrm{o} - \bar{\bm y}^\mathrm{f}), \\
    \delta \bm W^\mathrm{a} &= [(M-1) \tilde{\bm P}^\mathrm{a}]^{1/2}.
\end{eqnarray}

\noindent
Here, $\bm R$ is the matrix called the observation error covariance, which is a model constant determined by the observation noise.
$\tilde{\bm P}^\mathrm{a}$ is the analysis covariance in the ensemble space, given by

\begin{eqnarray}
    \label{eq:letkf_analysis_p}
    \tilde{\bm P}^\mathrm{a} = [
        (M-1)\bm I + \delta \bm Y^{\mathrm{f}\top} \bm R^{-1} \delta\bm Y^{\mathrm{f}}
        ]^{-1}.
\end{eqnarray}

Finally, the analysis state vector in the ensemble space is transformed back to the model space, so that the analysis state vector of the $m$-th ensemble member $\bm x_m^{\mathrm{a}}$ is computed as

\begin{eqnarray}
    \bm x_m^{\mathrm{a}} = \bar{\bm x}^\mathrm{f} + \delta \bm X^\mathrm{f} \left(
        \bar{\bm w}^\mathrm{a} + \delta \bm W^{\mathrm{a}}|_m
        \right),
\end{eqnarray}

\noindent
where $\delta \bm W^{\mathrm{a}}|_m$ denotes the $m$-th column of $\delta \bm W^\mathrm{a}$.

\subsubsection{Covariance localization and inflation}

In most practical applications, the EnKF faces a lack of ensemble size.
The ensemble size and the DOFs of the system are typically $M = \mathcal{O}(10)\sim \mathcal{O}(10^3)$ and $N = \mathcal{O}(10^4)\sim \mathcal{O}(10^{10})$, respectively.
In such EnKF, spatial localization of the covariance (covariance localization) is generally considered to reduce the pseudo-correlations coming from the sampling error due to the small ensemble size.
In this study, we utilize the covariance localization in the observation space based on the $\bm R$-\textit{localization}~\citep{Hunt2007a,Miyoshi2011}.
The $\bm R$-localization increases the observation error at the spatially distant observation points so that the influences from those points are weakened.
The localized covariance is given by
$\bm R_\mathrm{loc} = \bm G^{-1} \circ \bm R$.
Here, $\bm G$ is the matrix of localization weight, whose elements have values between 0 to 1, and the values generally decay depending on the spatial distance between the grid point, $\bm\xi$, and the observation points, namely, $\bm\eta$.
The symbol $\circ$ denotes the elementwise matrix-matrix multiplication (i.e., the Schur product).
In the case that the observation error of each variable is uncorrelated, $\bm R$ and $\bm G$ are diagonal, and the above equation can be rewritten as
$\bm R_\mathrm{loc}^{-1} = \bm G \circ \bm R^{-1}$.
Here, we employ the Gaspari-Cohn function as the definition of each element of $\bm G$.
This function gives a nearly Gaussian decay,
$e^{-1.5d^2}$,
approximated by

\begin{eqnarray}
    G(d) &=
    \left\{\begin{array}{ll}
        -\frac{1}{4}d^5 + \frac{1}{2}d^4 + \frac{5}{8}d^3 - \frac{5}{3}d^2 + 1, & 0\leq d \leq1 \\
        \frac{1}{12}d^5 - \frac{1}{2}d^4 + \frac{5}{8}d^3 + \frac{5}{3}d^2 - 5d + 4 - \frac{2}{3}d^{-1}, & 1\leq d\leq2 \\
        0, & d\geq 2,
    \end{array}\right.,
\end{eqnarray}

\noindent
where $d = |\bm\xi - \bm\eta|/d_\mathrm{cut}$ and $d_\mathrm{cut}$ is the cutoff parameter.
The merit to use this function is that the distant observation points can be explicitly ignored because the approximated value is exactly zero for $d\geq2$
(see \figref{fig:rloc}).
By using $\bm R$-localization, the observation error covariance $\bm R_\mathrm{loc}$ and thus the analysis vectors $\bar{\bm w}^\mathrm{a}, \delta \bm W^\mathrm{a}|_m$ have different values on each grid point. It leads to the batched computation of matrices, where each batch computes the analysis at each grid point. Such batched computation was parallelized and accelerated by GPU in our previous study~\citep{Hasegawa2022}.

It is also commonly known that the EnKF faces the so-called \textit{filter divergence} problem, in which too small error covariances result in spuriously tiny Kalman gains.
As a workaround for the filter divergence, we employ the multiplicative covariance inflation method~\citep{Anderson1999a}, namely,
$\bm P^\mathrm{f}_\mathrm{inf} = \beta \bm P^\mathrm{f}$,
where $\beta$ is the inflation coefficient.
The value of $\beta$ is, practically, hand-tuned \textit{ad hoc}~\citep{Hunt2007a,Miyoshi2007} or automatically estimated by the adaptive inflation algorithm~\citep{Anderson2007a,Miyoshi2011}.
In this study, the hand-tuned constant parameters were used.
It is noted that the hand-tuned inflation coefficients tended to have larger values with smaller ensemble sizes or with fewer observation points.

In the LETKF, the $\bm R$-localization and the multiplicative covariance inflation affect only $\bar{\bm w}^\mathrm{a}$ and $\tilde{\bm P}^\mathrm{a}$, and thus, Eqs.
\eqref{eq:letkf_barw} and \eqref{eq:letkf_analysis_p} are rewritten as

\begin{eqnarray}
    \bar{\bm w}^\mathrm{a} &= \tilde{\bm P}^\mathrm{a} \, \delta \bm Y^\mathrm{f\top} \bm R_\mathrm{loc}^{-1} 
        \left( \bm y^\mathrm{o} - \bar{\bm y}^\mathrm{f} \right),\\
    \tilde{\bm P}^\mathrm{a} &= \left(
        \frac{M-1}{\beta}\bm I + \delta \bm Y^{\mathrm{f}\top} \bm R_\mathrm{loc}^{-1} \delta\bm Y^{\mathrm{f}}
        \right)^{-1}.
\end{eqnarray}

\section{Experiment with 2D forced isotropic turbulence}
\label{s:exp}

\subsection{Design of numerical configuration}

\begin{table}
    \centering
    \caption{Detailed configuration of 2D forced isotropic turbulence.
        \label{tab:configure}
    }
    \begin{tabular}{lll}
     \hline
     symbol & description & value \\
     \hline
     $N$ & grid size & $256^2$ \\
     $u_0$ & reference velocity (Courant number) & 1  \\
     $\Delta t$ & time-stepping width & 0.001232 \\
     $\rho_0$ & reference fluid density & 1 \\
     $\nu$ & kinetic viscosity & $10^{-4}$ \\
     $\mu$ & friction force coefficient & $5\times10^{-4}$ \\
     $k_\mathrm{f}$ & center of energy injection wavenumber regime & 4 \\
     $\Delta k$ & width of energy injection wavenumber regime & 2 \\
     \hline
\end{tabular}
\end{table}

To clarify the properties of the ensemble DA, we need a numerical experiment, which shows chaotic dynamics and requires the minimum computational cost. 
The latter feature is essential to study the properties of the LETKF with various observation conditions up to large ensemble sizes.
For this purpose, we employ a minimal 2D problem of the forced isotropic turbulence, which is well understood by the classical theory by \citet{Kraichnan1967,Leith1968,Batchelor1969-hyphen}~(KLB theory).

We consider the 2D squared computational domain on the $x-y$ plane with the system size of $2\pi\times2\pi$, double periodic boundaries, and the grid number of $256\times256$.
The acceleration is given by the superposition of the energy injection and the friction,
$\bm F = \bm F^i + \bm F^\mu$.
Here, the energy injection $\bm F^i$ is given by

\begin{eqnarray}
    \bm F^i&=\Delta x\nabla F_e\times \bm i_z,\\ 
    F_e &= F_{e,0} \left(
    \sum_{||\bm k| - k_\mathrm{f}| \leq \Delta k}
      \bm r_1 \cdot \bm k \cos{(\bm k \cdot \bm \xi)} 
   + \bm r_2 \cdot \bm k \sin{(\bm k \cdot \bm \xi)}
   \right),
\end{eqnarray}

\noindent
where $\bm k$ is the wavenumber vector,
$\bm i_z$ is the unit vector in the direction of $z$-axis, 
$k_\mathrm{f}$ and $\Delta k$ are respectively the center and width of the energy injection wavenumber regime,
$\bm r_1$ and $\bm r_2$ are Gaussian random vectors with zero-mean and the standard deviation being $\Delta x$,
and $F_{e,0}$ is the energy injection parameter.
$F_{e,0}$ is determined so that the velocity becomes of the order of the reference velocity on average.
The friction is defined as

\begin{equation}
    \bm F^\mu = -\mu\bm u,
\end{equation}

\noindent
with the friction parameter $\mu$.
Here, the motivation to introduce the friction force is to avoid the immoderate energy increase in the low-wavenumber regime, which is known as the energy condensation phenomenon~\citep{Smith1994,Chertkov2007}.
This is essential for quasi-steady long-term simulations.
The details of other parameters are summarized in \tabref{tab:configure}, where the normalization is chosen so that the reference velocity $u_0$ and the reference density $\rho_0$ are unity.
The time-stepping width $\Delta t$ is adjusted so that the Courant number to the reference velocity is set to 0.05.
In these parameters, the SGS viscosity $\nu_\mathrm{SGS}$ has the value of the same order as the kinetic viscosity $\nu$.

\begin{figure}
    \centering
    \includegraphics[width=.8\linewidth]{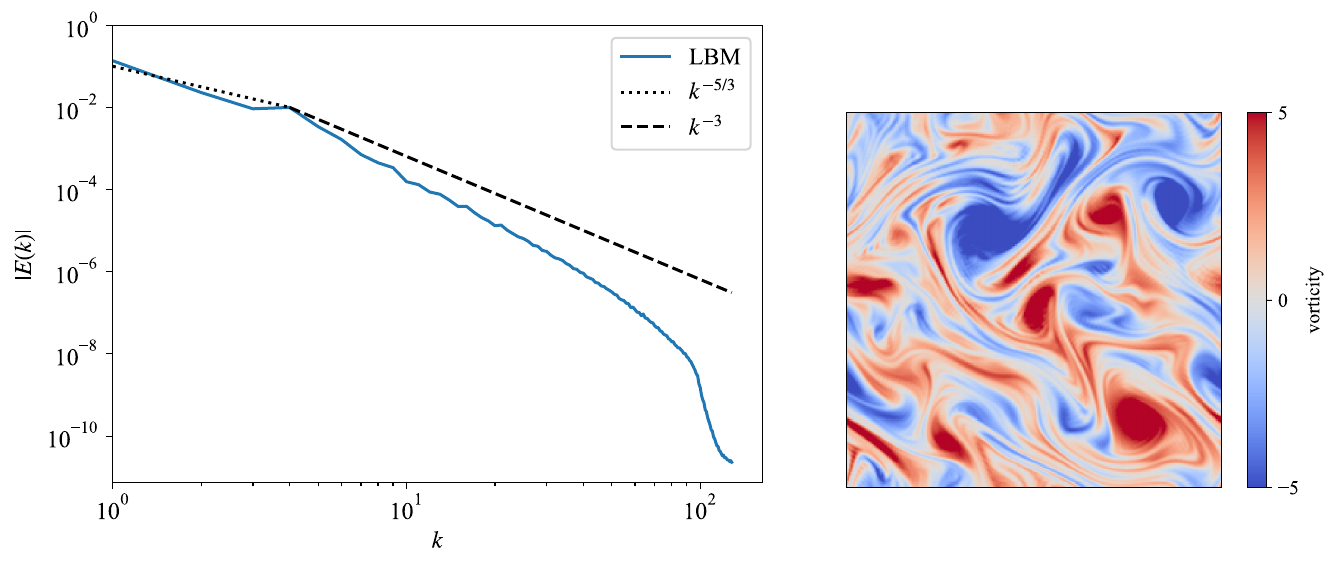}
    \caption{Typical energy spectrum and vorticity map.}
    \label{fig:naturerun}
\end{figure}

\figref{fig:naturerun} shows a typical quasi-stationary state of the computed 2D turbulence.
Here, the energy spectrum is defined by the shell average over the 2D wavenumber:
\begin{eqnarray}
    E(k) &= \sum_{||\bm k'| - k|\leq1/2} E(\bm k'), \\
    E(\bm k) &= \frac{1}{2}\ \left|\hat{\bm u}(\bm k)\right|^2,
\end{eqnarray}
where $\bm k$ is the 2D wavenumber, $k$ is the 1D wavenumber for the shell average,
$\hat{\bm u}(\bm k)$ is the Fourier transform of $\bm u$.
There exists the energy spectrum with the dual cascade, composed of the inverse cascade for $k < k_\mathrm{f}$, and the normal cascade for $k_\mathrm{f} < k < k_\mathrm{d}$.
The dissipation regime is also observed in $k > k_\mathrm{d}$, where $k_\mathrm{d} \approx 90$.
Although the dual cascade was observed,
the normal cascade does not follow the power law of $k^{-3}$ and is characterized by the steeper slope $\approx k^{-4}$.
Although it seems to be inconsistent with the KLB theory, it is a well-known nature in isotropic turbulence simulations.
The KLB theory assumes the limit of infinite resolution and/or Reynolds number, which cannot be satisfied in simulations with finite resolution, and thus, viscosity.
Theoretically, the power law for the normal cascade is modified to the range between $k^{-3}$ and $k^{-5}$ for the moderate resolution and Reynolds number~\citep{Tran2002,Tran2003}.
The power lows of the normal cascade in former numerical studies were also steeper than $k^{-3}$~\citep{Herring1985,Legras1988,Scott2007,Tsang2009,Clark2020,Xie2020}.
It is noted that simulations at higher resolution may make the result close to the KLB theory (cf. the work by \citet{Boffetta2007}: the mesh resolution up to $32,768^2$ resulted in the power low of $\approx k^{-3.35}$);
however, such a larger scale simulation is not in the scope of the current study, i.e., the minimal benchmark of the ensemble DA with the LBM-LETKF.
In addition, the LBM solves weakly-compressible flows rather than incompressible flows, and thus, flow characteristics may be slightly differ from the KLB theory.
For these reasons, in this study, we accept the normal cascade with the power law steeper than $k^{-3}$ and adopt this numerical setup as the minimal benchmark problem.

\subsection{Lyapunov exponent of the turbulence}
\label{ss:lyap}

\begin{figure}
    \centering
    \includegraphics[width=.8\linewidth]{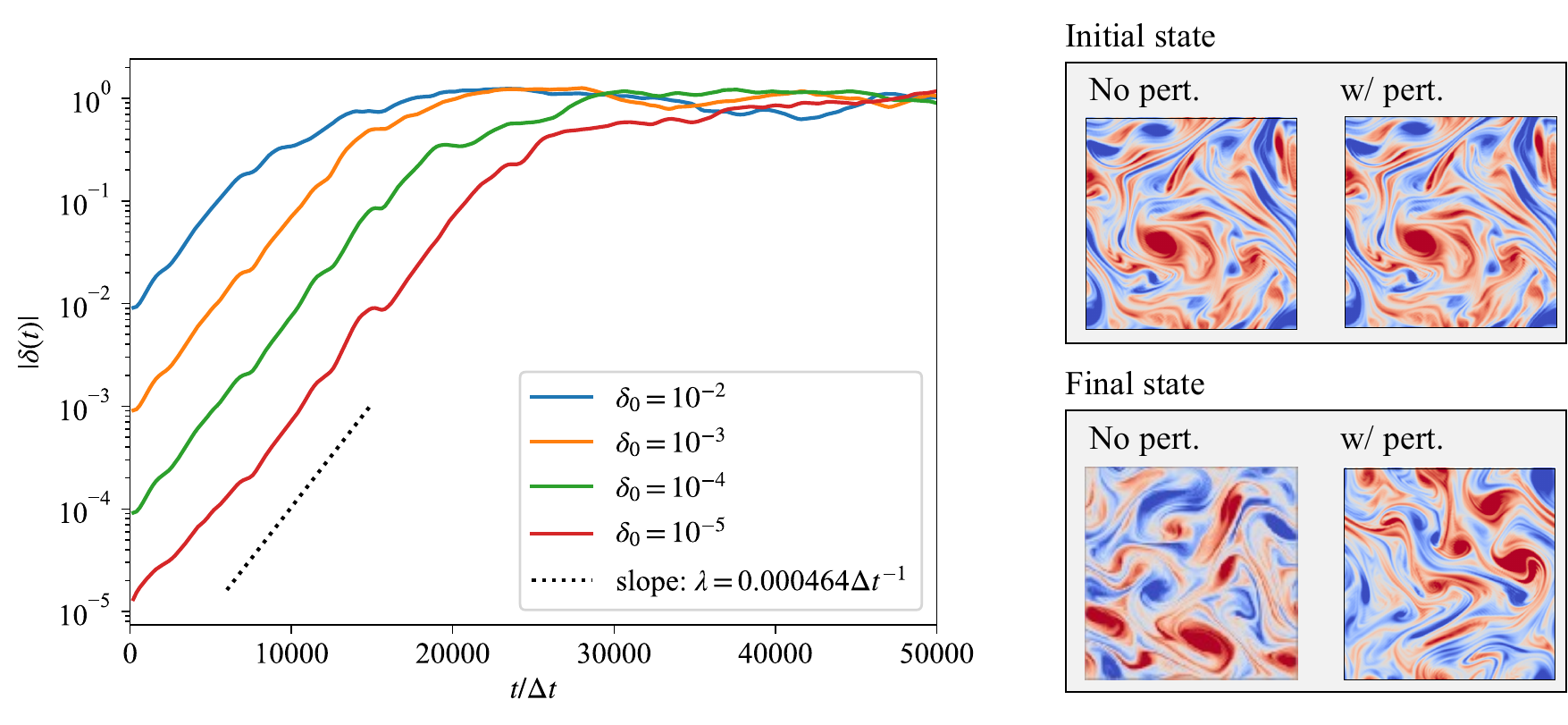}
    \caption{Error growths with various magnitudes of the initial perturbations.
        (Left) time history of the errors.
        (Right) vorticity maps at the initial and final states with/without the perturbation ($\delta_0=10^{-5}$).
        \label{fig:lyap}
    }
\end{figure}

Firstly, we investigated the error growth (also known as the uncertainty of the initial condition) at the designed numerical setup.
The purpose is to estimate the maximum Lyapunov exponent, which indicates the predictability of turbulence~\citep{Nastac2017} and gives guidance to design the time interval of the continuous DA~\citep{Labahn2020}.
In this test, we prepared the following perturbed distribution function $f^\mathrm{pert}_\alpha$ at $t=0$,

\begin{eqnarray}
    f_\alpha^\mathrm{pert} &= f_{\alpha,0} + \delta_0 \left(
    f_\alpha^\mathrm{eq}(\rho_0,u^\mathrm{rand},v^\mathrm{rand}) - f_{\alpha,0}\right), 
\end{eqnarray}

\noindent
where $u^\mathrm{rand}, v^\mathrm{rand}$ are given by Gaussian random numbers with zero-mean and the standard deviation being the reference velocity $u_0$.
$f_\alpha^\mathrm{eq}(\rho_0,u^\mathrm{rand},v^\mathrm{rand})$ is the equilibrium distribution function with the constant density $\rho_0$ and the random velocity ($u^\mathrm{rand}, v^\mathrm{rand}$),
and $\delta_0$ is the perturbation magnitude.
Here, the time $t=0$ is not the initial step of the simulation;
the original distribution function $f_{\alpha,0}$ is prepared by the spin-up simulation for $10^6\Delta t$, during which the system is fully-developed and is converged to the quasi-stationary state.

The separation of the simulations with and without the perturbation was evaluated by

\begin{equation}
    |\delta(t)| = \sqrt{
        \frac{1}{N}
        \sum_{\bm\xi}\left|\bm u^\mathrm{pert}(\bm \xi, t) - \bm u(\bm \xi, t)\right|^2
    },
\end{equation}

\noindent
where $\bm u^\mathrm{pert}$ and $\bm u$ are the velocities in perturbed and unperturbed simulations, respectively.

\noindent
Following the chaotic nature of turbulence, the error grows exponentially; then, $|\delta(t)|$ may be approximated by

\begin{equation}
    |\delta(t)| = e^{\lambda t} |\delta(0)|,
\end{equation}

\noindent
where $\lambda$ is the error growth rate, namely, the maximum Lyapunov exponent.

We evaluated the maximum Lyapunov exponent against the various ranges of the initial perturbations from $10^{-5}$ to $10^{-2}$.
In \figref{fig:lyap}, the time history of the error and the vorticity contour map indicate the growth of the error.
As seen in the vorticity contour map, the initial states are quite similar since the initial perturbation is small enough compared to the turbulence;
in contrast, the final states are completely different between perturbed and unperturbed simulations,
which reflect the error growth following the chaotic dynamics of turbulence.
The error grows exponentially and saturates at the level of $\delta(t)\sim\mathcal{O}(1)$ as seen in the time history.
The maximum Lyapunov exponents, corresponding to the slopes of the time histories, were almost the same regardless of the magnitude of the initial perturbation.
The value of the maximum Lyapunov exponent was $\lambda\sim4.64\times10^{-4}\Delta t^{-1}$, and then, the predictability time was estimated as $t_\mathrm{pred} = \lambda^{-1} \sim 2,150\Delta t$.

The time interval of the continuous DA for chaotic systems has to be frequent enough compared to the predictability time~\citep{Labahn2020}.
Hence, we determined the design value of the sub-timesteps per DA cycle as $s = 200$, which refers to $s\Delta t \lessapprox t_\mathrm{pred}/10$.

\subsection{Data assimilation experiments}

\begin{figure}
    \centering
    \includegraphics[scale=0.5]{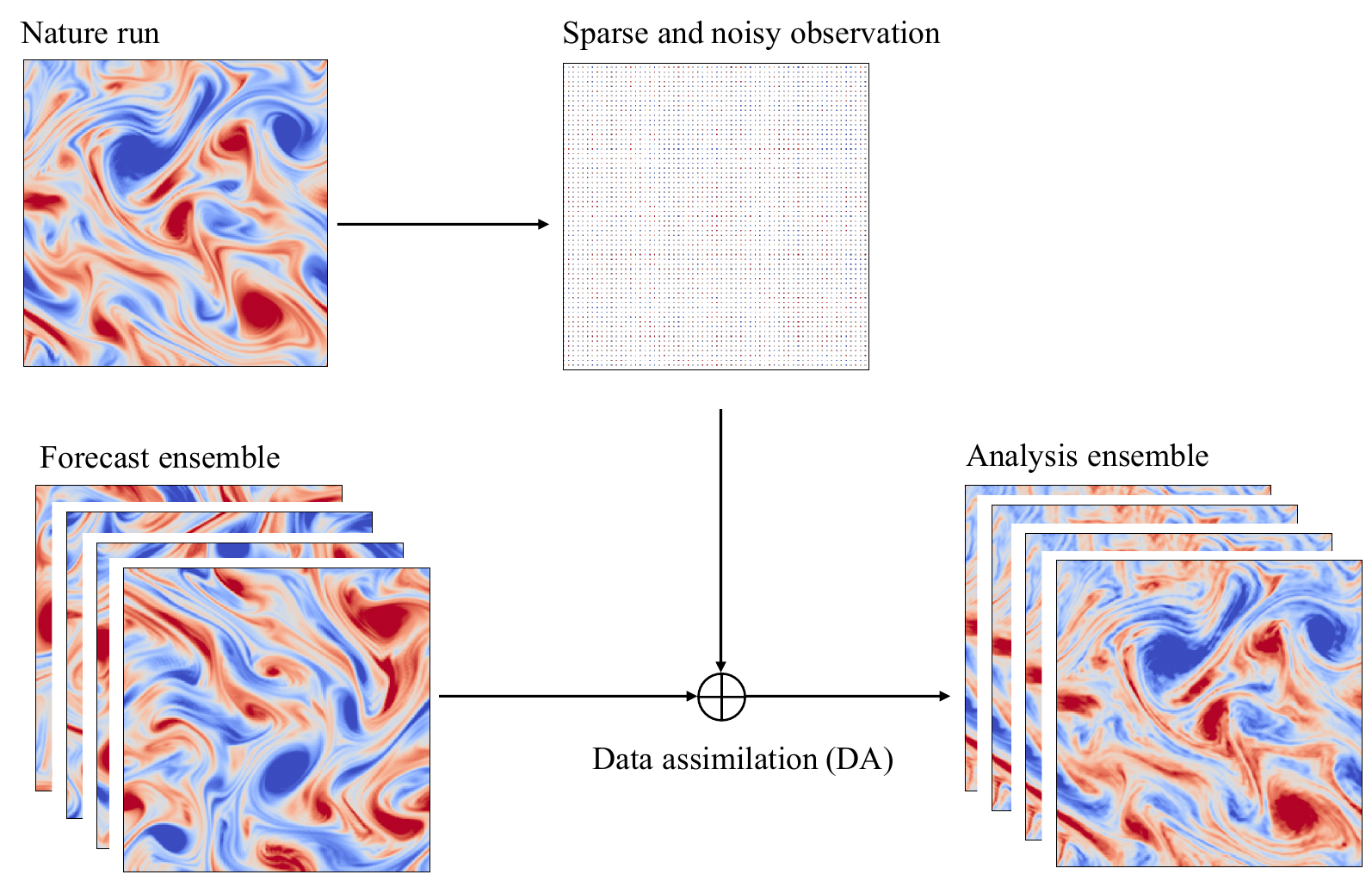}
    \caption{
    Schematic illustration of the observing system simulation experiment (OSSE) for the 2D turbulence simulation.
    (Top left) Ground truth.
    (Bottom left) \textit{Forecast} ensemble, which is the set of trial runs to reproduce the ground truth.
    Since the ground truth is invisible and the system is chaotic, the forecast ensemble may result in different states.
    (Top middle) Observation, which is created from the ground truth with artificial observation noises.
    (Bottom right) the data-assimilated state or the analysis ensemble, that is generated using the observation data to reproduce the ground truth.
    }
    \label{fig:osse_concept}
\end{figure}

\begin{figure}
    \centering
    \includegraphics[width=.35\linewidth]{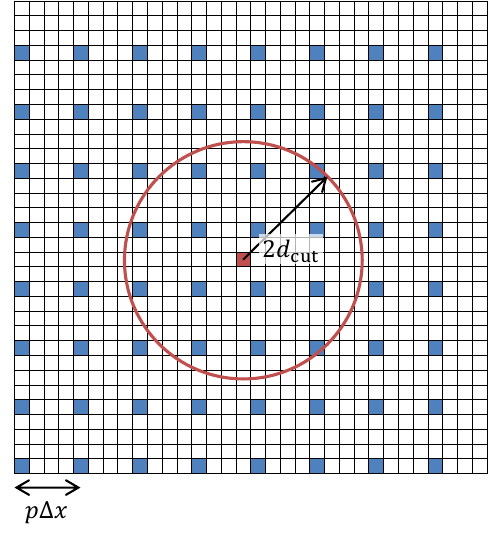}
    \caption{An illustration of the spatially-sparse observation and the kernel of $\bm R$-localization. 
    Each cell is the computational grid, while
    the blue and the white cells are respectively the grid with and without the observation.
    The red cell depicts the target grid assimilated with observation points within the kernel shown by the red circle.
    $p$ is the observation interval ($p=4$ in this example).
    $d_\mathrm{cut} = p\Delta x$ is the cutoff parameter for the $\bm R$-localization.}
    \label{fig:rloc}
\end{figure}

As shown in \figref{fig:osse_concept}, we carried out the DA experiment in the observing system simulation experiment (OSSE) manner.
First, the ground truth was computed with a certain initial condition.
To emulate the noisy observation, the macroscopic observable variables
$(u^\mathrm{o}, v^\mathrm{o}, \rho^\mathrm{o})$ 
were measured with artificial noises on equidistant sampling points in the ground truth.
Here, the density $\rho^\mathrm{o}$ was also measured because the density in the LBM reflects the pressure ($P = \rho/3$), which would be also observable in real experiments.
The artificial observation noises were given by Gaussian random numbers with zero-mean, whose standard deviations for $u^\mathrm{o}$/$v^\mathrm{o}$ and $\rho^\mathrm{o}$ were 10\% of $u_0$ and 1\% of $\rho_0$, respectively.
Here, the noise amplitude for the density differs from that for the velocity because the density corresponds to the pressure.
The noise amplitude for the density is designed based on its fluctuation $<\pm5\%$. 
The spatiotemporal resolution of the observation was designed as follows:
\begin{itemize}
    \item 
    The observation points were sampled uniformly with the spatial interval $p$, which gave $256/p\times256/p$ observation points.
    The observation condition was varied from the dense observation case, $p=1$, to the sparse observation cases, $p=2, 4, \dots, 64$.
    \item
    The temporal interval of the observation was chosen to be the same as the interval of the continuous DA, $s=200$.
\end{itemize}
Next, the other simulations were prepared with the random initial conditions.
The initial conditions for the ground truth and the ensemble simulations are all uncorrelated with each other. 
We conducted the simulations without the DA, with the nudging, and with the LETKF.
In the nudging cases, the gain parameter was tuned by the preliminary numerical experiments. The optimal parameters were $\gamma=0.11$ in the dense observation case (\sref{ss:da-dense}) and $0.11\leq\gamma\leq0.31$ in the sparse observation cases (\sref{ss:da-sparse}). Here, more sparse observation leads to larger $\gamma$.
In the LETKF cases, the ensemble sizes were chosen as $M = 4, 16, 64$ and the inflation parameter was optimized in each case.
In the $\bm R$-localization, the cutoff parameter was chosen to be $d_\mathrm{cut} = p\Delta x$, which guarantees that each computational grid is affected by $\sim {\cal O}(10)$ observation points within the distance of $2d_\mathrm{cut}$ (see \figref{fig:rloc}).

\subsubsection{Accuracy evaluation at the dense observation case\label{ss:da-dense}}

\begin{figure}
    \centering
    \includegraphics[width=.5\linewidth]{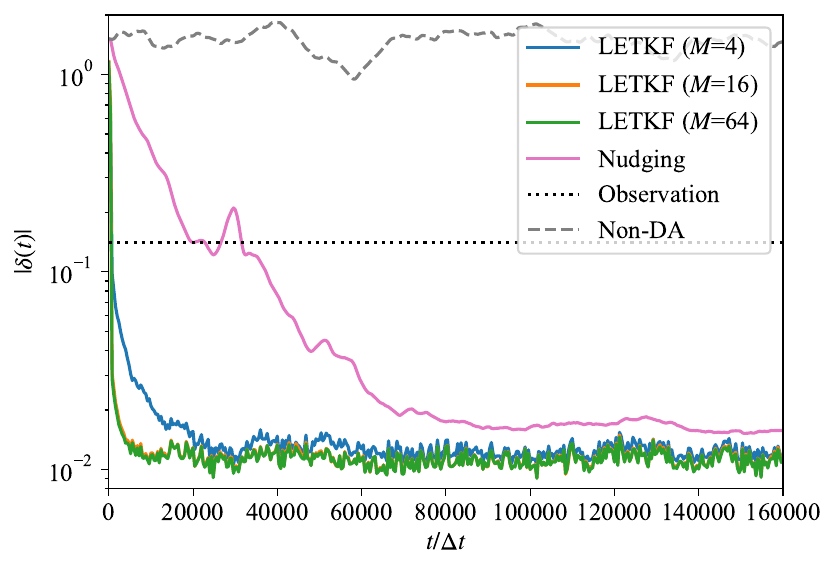}
    \caption{
    Time histories of the RMSEs in the OSSEs without the DA, with the nudging, and with the LETKF at the observation interval $p=1$.
    The black dotted line $\delta=0.142$ depicts the root-mean-square (RMS) of the observation noise.
    \label{fig:timehist}
    }
\end{figure}

\begin{figure}
    \centering
    \includegraphics[width=.5\linewidth]{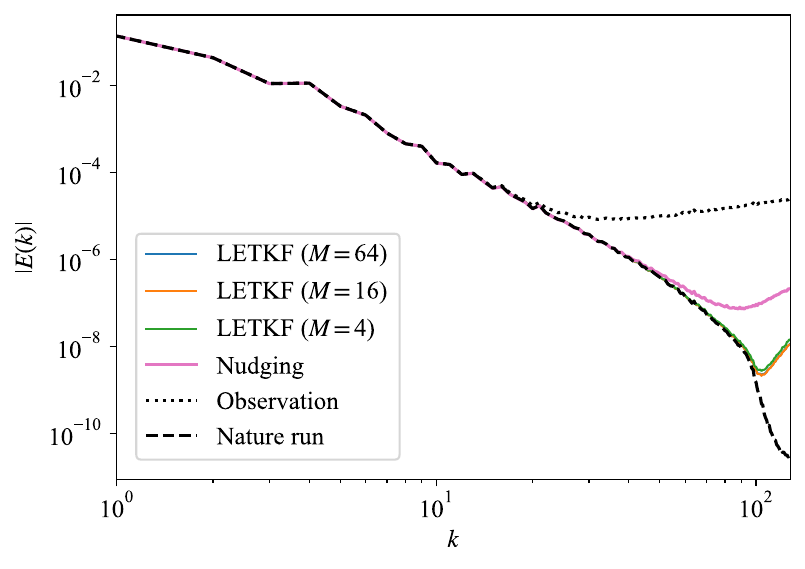}
    \caption{Energy spectra in the OSSEs with the nudging and with the LETKF at the observation interval $p=1$. Also shown are the energy spectra of the ground truth and its observation data with artificial noises.
        \label{fig:fft_spectrum_p1}
    }
\end{figure}

Firstly, we investigated the DA experiment with the dense observation case, $p=1$.
The DA accuracy was defined by the root means squared error (RMSE), namely,

\begin{equation}
    |\delta(t)| = \sqrt{\frac{1}{N}\sum_{\bm\xi} |{\bm u}^\mathrm{DA}(\bm\xi, t) - \bm u^\mathrm{t}(\bm\xi, t)|^2},
\end{equation}

\noindent
where $\bm u^\mathrm{t}$ is the velocity in the ground truth state, and $\bm u^\mathrm{DA}$ is the velocity obtained by each DA approach.
In the LETKF, the ensemble mean, ${\bm u}^\mathrm{DA} = M^{-1}\sum_{m=0}^{M-1} \bm u_m$, is considered as the best estimate.
\figref{fig:timehist} shows the time history of the RMSE for each DA approach.
The errors in all cases were $\mathcal{O}(1)$ at $t=0$ and became smaller in time except for the non-DA case.
They converged into the values of 
0.0190 (nudging),
0.0121 (LETKF, $M=4$),
0.0110 (LETKF, $M=16$), and
0.0109 (LETKF, $M=64$),
while the RMSE of the observation noise was 0.142.
It is noted that the observation noise was $0.1$ for each of $u^\mathrm{o}, v^\mathrm{o}$, thus, the observation noise of the velocity magnitude was $0.1\times\sqrt2=0.142$.
The converged value is calculated by the time-averaging over the converged state for $70,000\leq t/\Delta t\leq80,000$.
Although both the nudging and LETKF cases achieved good results, the LETKF cases showed lower RMSEs than the nudging case.
The LETKF gave better results at larger ensemble sizes, and its accuracy was almost converged to the best value for $M\geq 16$.
Besides, the duration to converge the RMSE was shorter in the LETKF cases than in the nudging cases.
In the LETKF cases, the RMSE reached around the lower bound at $t/\Delta t=10,000\sim20,000$, while the nudging case required $\sim70,000\Delta t$ to reach the converged state.
This feature is due to the difference in the gain:
In the nudging, the gain parameter is a constant value tuned for the fully assimilated state ($t/\Delta t\gtrapprox 70,000$);
thus, the gain parameter might be too small for the initial transient period ($t/\Delta t < 70,000$), where the calculation error was larger than the observation error.
In contrast, in the LETKF, the gain is dynamically determined by the error covariances of the state vector and the observation vector, which result in the larger gain at the larger error covariance during the initial transient period.
This feature of the LETKF would be convenient to reduce the spin-up time from the non-assimilated state to the fully assimilated state.
Here, it should be noted that the faster convergence may be achieved also in the nudging if one introduces an adaptive optimization of $\gamma$, so that it becomes larger in the initial transient period.
However, it is hard to optimize $\gamma$ in an adaptive manner, because in reality, we cannot estimate the RMSE against the true state and the nudging does not have automatic optimization strategy to fit the model to the true state. 
On the other hand, the LETKF can automatically optimize the Kalman gain based on the ensemble statistics.
Therefore, the adaptive nudging might require large number of test runs to try different time-varying parameters by hand-tuning.
In addition, such a hand-tuned parameter might not be optimal for other initial conditions.
Thus, we did not consider to introduce the adaptive nudging in this study.

To see the details of the fully assimilated state in the nudging and LETKF cases, we examined the energy spectra.
In the LETKF, the energy spectrum was evaluated by the mean of the energy spectra in each ensemble member $|E_m(k)|$.
\figref{fig:fft_spectrum_p1} shows the energy spectra which are averaged over $t/\Delta t=76,000\sim 80,000$.
Since the observation error was given by a Gaussian random noise (i.e. a white noise), the energy spectra of the observation error were almost constant $\Delta E^\mathrm{o}(k) \sim O(10^{-4})$ as seen in the figure.
In the low wavenumber regime, the observation error was small enough, and both the nudging and LETKF cases showed good agreement with the ground truth.
The difference of the energy spectra can be seen only in the high wavenumber regime $k \gtrapprox 50$.
The result of the nudging case was closer to the observation spectrum and deviated from the ground truth;
in contrast, the LETKF case showed a good agreement with the ground truth up to the high wavenumber regime.
The visible error in the LETKF case was seen only in the dissipation regime $k \gtrapprox 90$.

These results might be caused due to the following reasons.
Firstly, the nudging potentially induced larger errors because:
\begin{itemize}
    \item A constant gain parameter $\gamma$ is assumed, and there is no mechanism to compensate for the influence of the observation error.
    \item There is no exact transform from the macroscopic variables $(u^\mathrm{o}, v^\mathrm{o}, \rho^\mathrm{o})$ to the velocity distribution function $f_{\alpha}^\mathrm{o}$.
    In the nudging case, the equilibrium distribution $f_{\alpha}^\mathrm{(eq,o)}$ was employed for reconstructing a reference state vector, which might induce extra errors.
\end{itemize}
Secondly, in the LETKF, these issues were properly treated:
\begin{itemize}
    \item The Kalman gain is optimized based on the ensemble statistics by taking account of the observation error, which improves the accuracy much better than the constant gain in the nudging.
    \item The observation data from neighboring observation points are considered, leading to a non-local DA process.
    This contains much more information than nudging based on the observation data from a single observation point.
\end{itemize}
Therefore, in applying the DA to the LBM, the LETKF is more straightforward and accurate than nudging.

\subsubsection{Accuracy evaluation at the sparse observation cases\label{ss:da-sparse}}

\begin{figure}
    \centering
    \includegraphics[width=.5\linewidth]{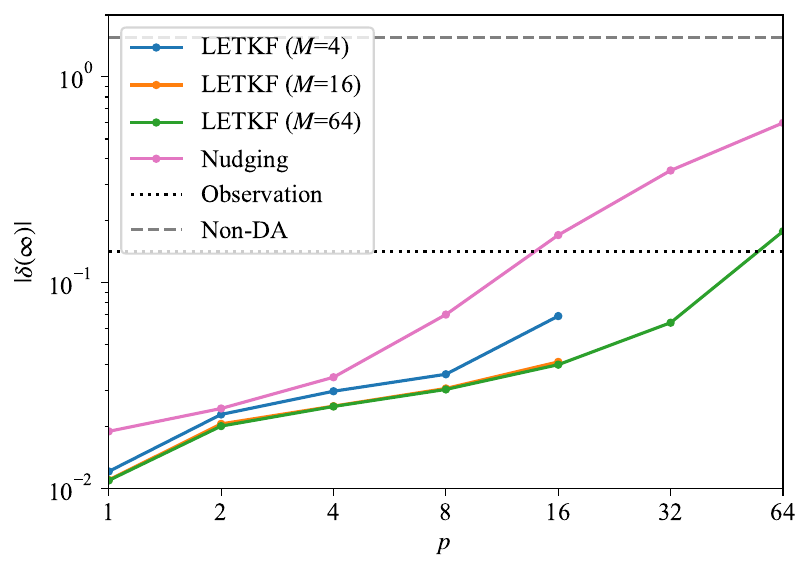}
    \caption{
        Dependencies of the RMSEs on the interval $p$ of spatially-sparse observation points.
        $|\delta(\infty)|$ refers to the time average of $|\delta(t)|$ over the converged state in $70,000\leq t/\Delta t \leq 80,000$.
        The black dotted line shows the RMS of the observation noise, $\varepsilon^\mathrm{o}=0.142$.
        Missing points in the LETKF indicate the occurrence of catastrophic filter divergence.
    }
    \label{fig:prune}
\end{figure}

\begin{figure}
    \centering
    \includegraphics[width=.49\linewidth]{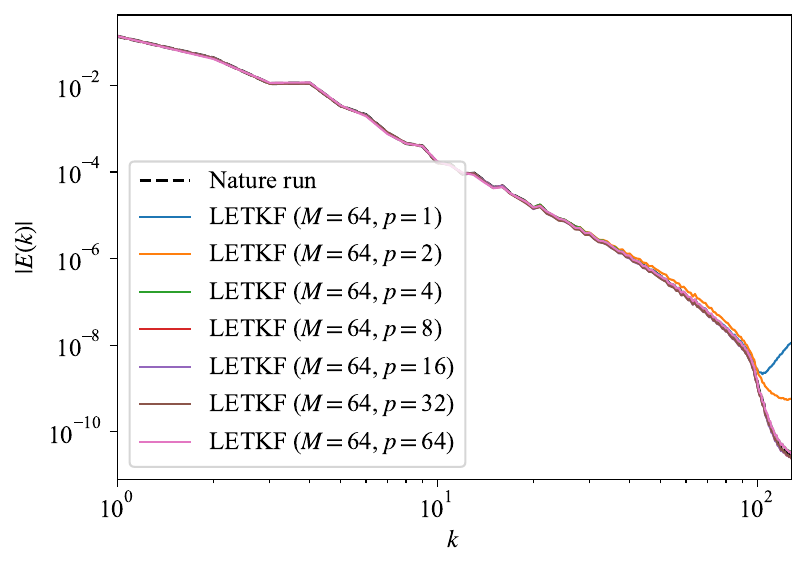}
    \includegraphics[width=.49\linewidth]{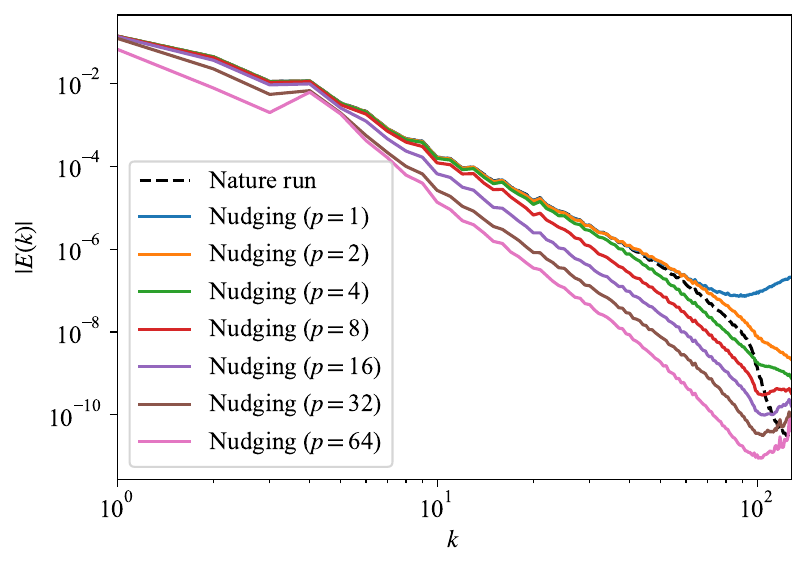}
    \caption{Energy spectra in the LETKF and the nudging with the various observation intervals, $p=1,2,4,\dots,64$. 
    }
    \label{fig:prune_fft}
\end{figure}

\begin{figure}
    \centering
    \includegraphics[width=.9\linewidth]{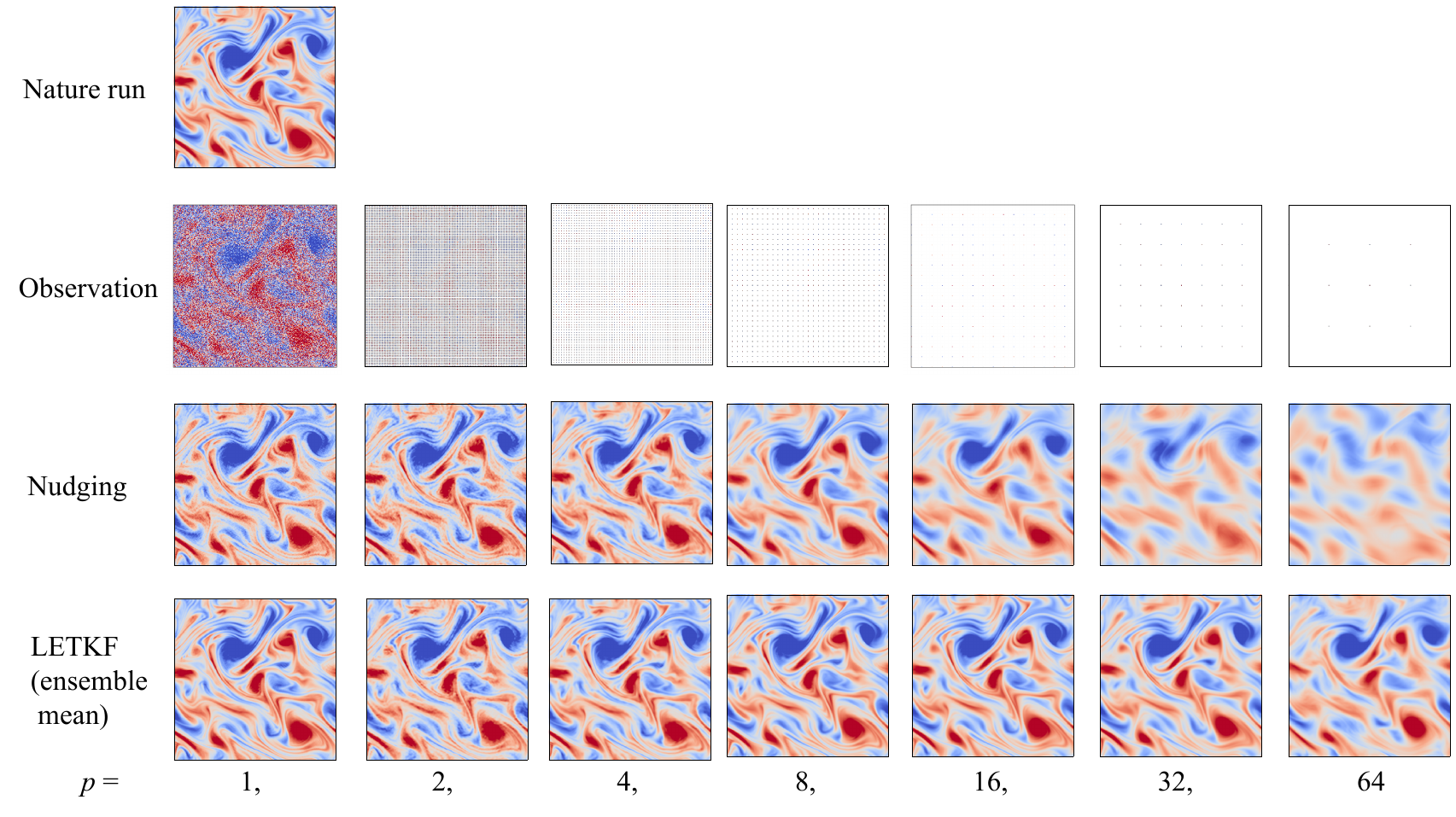}
    \caption{Vorticity maps in the DAs with the various observation intervals, $p=1,2,4,\dots,64$.
        \label{fig:prune_vor}
    }
\end{figure}

Next, we investigated the sparse observation cases.
The RMSEs observed in the experiments were summarized in \figref{fig:prune}.
The numerical experiments were conducted up to $p=64$, and points not shown in the figure mean numerically unstable cases.
This kind of numerical instability
was known as the catastrophic filter divergence problem~\citep{Harlim2010,Gottwald2013,Kelly2015},
where small ensemble sizes and sparse observation points produced poor statistics and uncorrelated observation data, leading to inaccurate estimations of the error covariance and thus the Kalman gain.
The LETKF cases with $M=4, p\geq32$ and with $M=16, p=64$ were unstable and diverged, while $M=64$ was stable in all the cases.
These results suggest that the shortage of the observation points could be avoided by increasing the ensemble size, and $M=64$ was sufficient for the current numerical experiments.
Here, it should be noted that the numerical instability of the EnKF in turbulent flows was discussed by \citet[Appendix D]{Suzuki2017}.
They proposed a relaxation factor to mitigate the improper injection energy at the initial few DA cycles, where the randomly distributed initial states of the ensemble give the large amplitude of the Kalman gain.
This initial numerical instability appears to be different from the catastrophic filter divergence.
In our case, however, the numerical instability occurred in the converged state where the ensemble variance and thus the background error were converged into small enough values.
Except for the numerical instability, the DA accuracy in the LETKF was always better than that in the nudging.
Here, we scored the DA accuracy against the observation noise $\epsilon^\mathrm{o}=0.142$, and the result was classified as \textit{good} when the RMSE was less than $\epsilon^\mathrm{o}$.
The nudging showed good results for $p\leq8$, while the LETKF showed good results up to $p = 32$.

To clarify the difference between the LETKF and nudging cases, we showed their energy spectra in \figref{fig:prune_fft} and vorticity maps in \figref{fig:prune_vor}.
In the nudging cases, the energy spectra with $p\ge 4$ indicated the normal cascade with a steeper power law than that in the ground truth, and this affected also the inverse cascade regime when the observation interval $p$ was larger.
This was attributed to the linear spatial interpolation of the observation data.
The underestimation of the energy spectra lead to unphysical vorticity maps, in which eddy structures were smeared out.
Once such unphysical flows were generated, the nudging could not recover the system in the correct direction.
In the nudging cases, the above issues produced a worse result than the above estimate.
In contrast, the LETKF cases showed better agreements with the ground truth. 
The energy spectra showed good agreement except for the energy dissipation regime.
The vorticity maps also indicated reasonable agreements, while only the small-scale eddies were smoothed out due to the ensemble average of vorticity fields.
In the current numerical experiments, the wavenumber regime of the energy injection due to the forced 2D turbulence was $k_\mathrm{f}=4$, and larger scale flows were developed via the inverse energy cascade. 
The ratio of the energy contained below a given wavenumber $k$ can be estimated by
\begin{equation}
    R(k) = 
    \frac{
    \sum_{k'=1}^{k} E^\mathrm{t}(k')
    }{
    \sum_{k'=1}^{k_\mathrm{max}} E^\mathrm{t}(k')
    },
\end{equation}
\noindent
where $k_\mathrm{max} = 128$ and $E^\mathrm{t}(k)$ denotes the 1D energy spectrum in the ground truth.
The value at the injection wavenumber was $R(k_\mathrm{f}) = 89.9\%$, therefore,
the DA accuracy was mainly determined by the inverse cascade regime.
According to the sampling theorem, the minimum required sampling interval for $k_\mathrm{f}=4$ is estimated as $p\leq k_\mathrm{max}/k_\mathrm{f} = 32$.
As seen in the vorticity maps in \figref{fig:prune_vor}, the results in the nudging were smeared out for $p\geq32$.
In the LETKF, such numerical diffusion of the vorticity map was largely suppressed, and the vortices at the energy containing-scale was well maintained up to $p=64$, which exceeded the above criterion.
This result suggests that the LETKF could assimilate the flow at finer scale than that of the observation.
This might be attributed to the covariance matrices, which takes account of the spatial correlation at finer scale.

Consequently, the DA experiment in this section revealed that the LETKF is highly robust and accurate than the nudging, in particular, at the spatially-sparse observation condition.

\subsubsection{Uncertainty and stability evaluation}

\begin{figure}
    \centering
    \includegraphics[width=.8\linewidth]{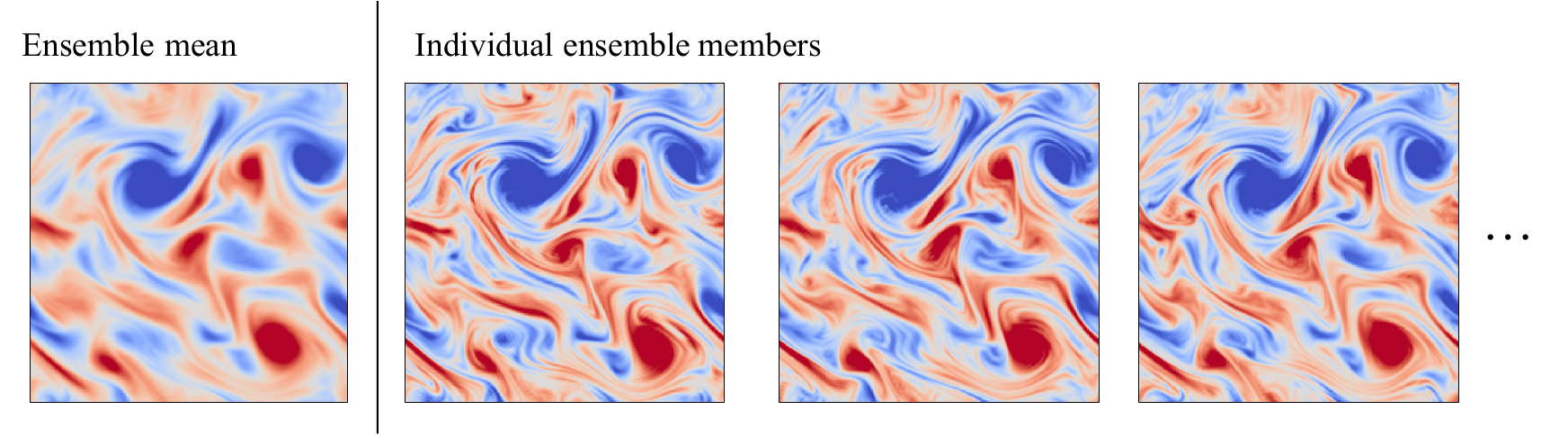}
    \caption{Vorticity maps for the ensemble mean and the individual ensemble members at the ensemble size of $M=64$ and the observation interval of $p=64$.
    The ensemble mean shows the same result as \figref{fig:prune_vor} (bottom, right).
    }
    \label{fig:vor_ind}
\end{figure}

\begin{figure}
    \centering
    \includegraphics[width=.6\linewidth]{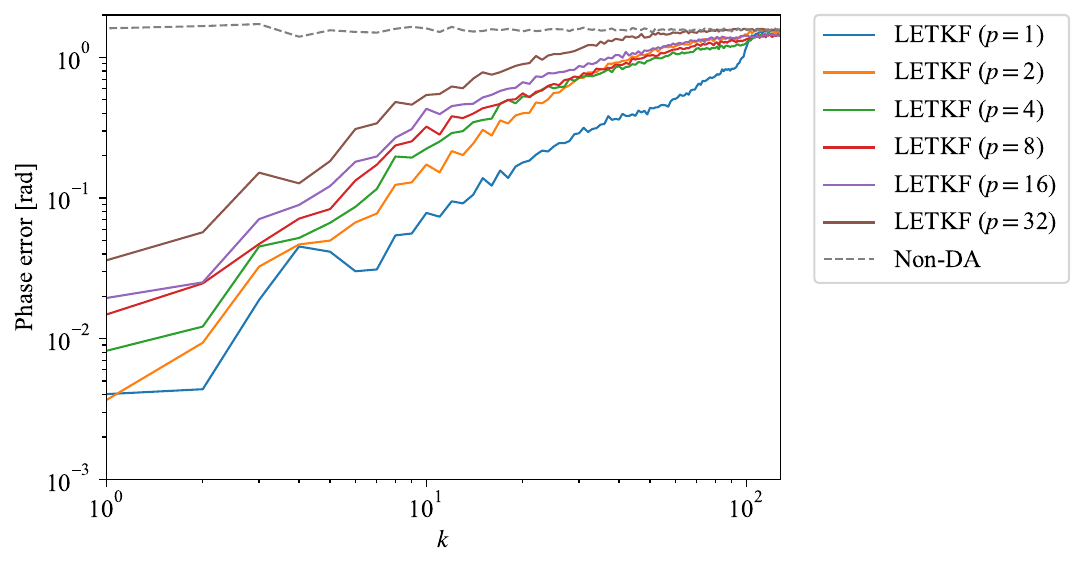}
    \caption{Phase errors of the energy spectra in the LETKF ($M=64$).
    \label{fig:letkf_fft_argerr}
    }
\end{figure}

\begin{figure}
    \centering
    \includegraphics[width=\linewidth]{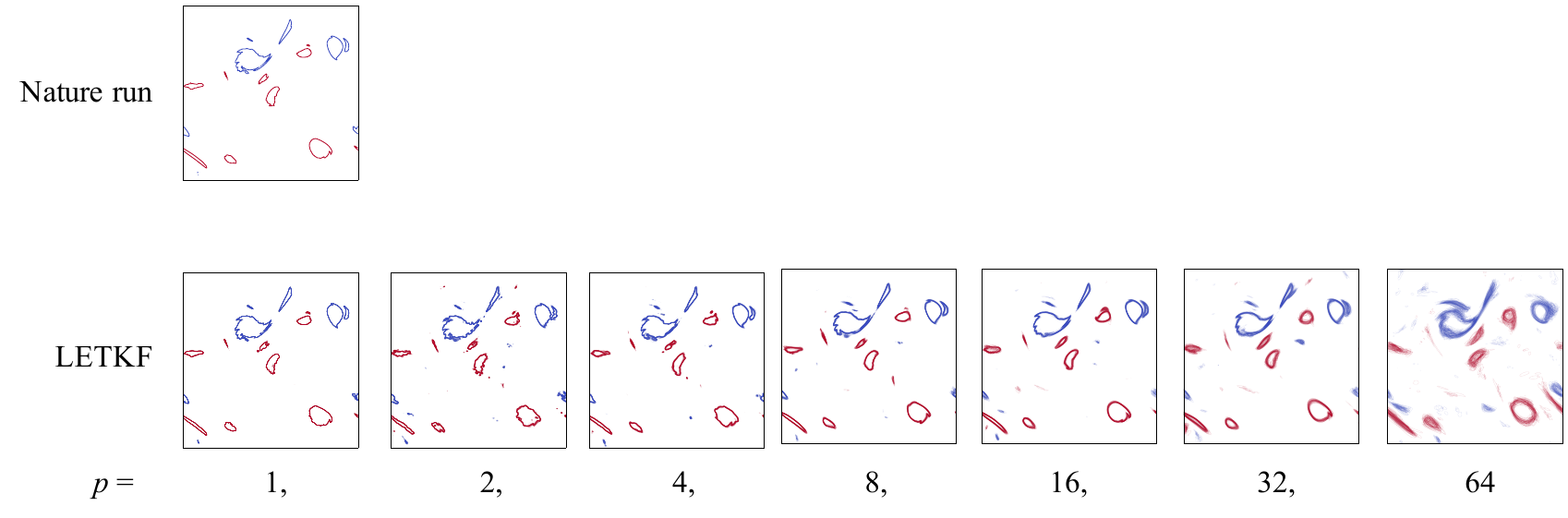}
    \caption{Spaghetti plots of the vorticity contours at the ensemble size $M=64$,
    where the isolines of the vorticity of $\omega=-5$ (blue) and $\omega=5$ (red) are plotted.
    \label{fig:spaghetti}
    }
\end{figure}

\begin{figure}
    \centering
    \includegraphics[width=.49\linewidth]{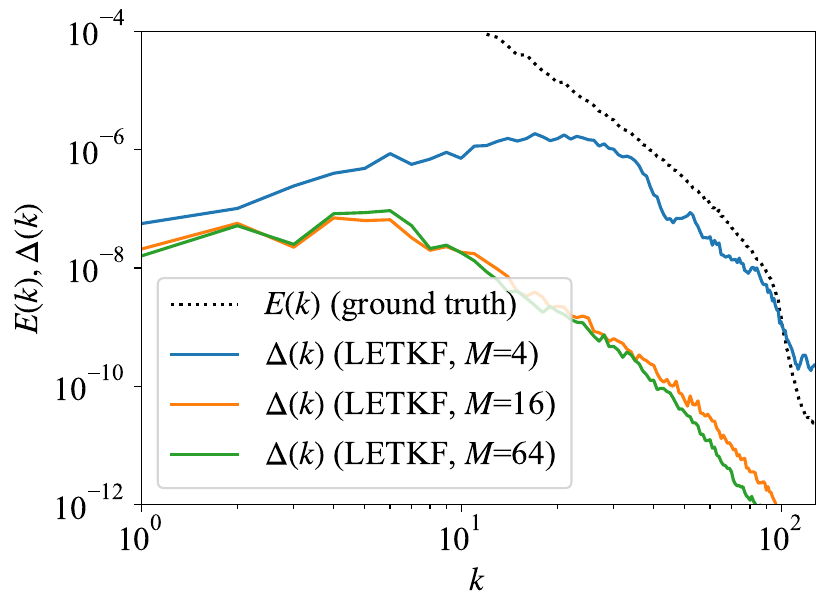}
    \includegraphics[width=.49\linewidth]{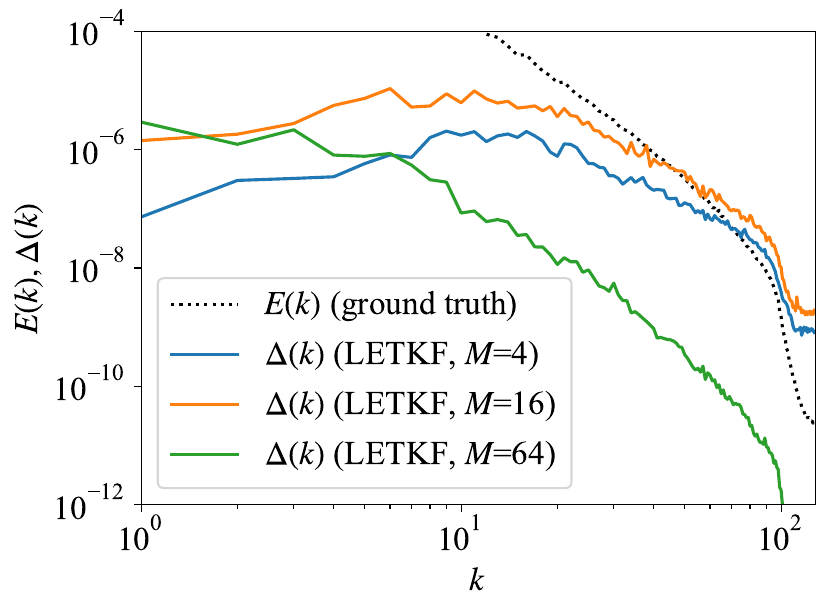}
    \caption{
    Energy spectra of the analysis increment $\Delta(k)$, compared with $E(k)$.
    Left shows the cases at $p = 32$, where $\Delta(k)$ was observed at $t/\Delta t = 5400$, just before the divergence of the case with $M=4$.
    Right shows the cases at $p = 64$, where $\Delta(k)$ was observed at $t/\Delta t = 3200$, just before the divergence of the cases with $M=4, 16$.
    \label{fig:letkf_z_matrix}
    }
\end{figure}

The previous section demonstrated the excellent DA capability of the LETKF against the nudging. 
In this section, we present further details of the ensemble statistics and their relation to the uncertainty and numerical instability in the LETKF.

\figref{fig:vor_ind} shows the ensemble-averaged vorticity and that in each ensemble member.
The ensemble average smeared out small-scale vortices in each ensemble member, while the individual members kept the small-scale eddies but their positions were slightly different from each other.
This indicates that the LETKF did not work as the numerical viscosity; in other words, the amplitudes of the energy spectra were well maintained up to the high wavenumber regime, while only the phase information was ruined due to the shortage of observation points.
The evidences are shown in \figref{fig:prune_fft} (left) and \figref{fig:letkf_fft_argerr}:
The former shows that the amplitudes of the energy spectra agreed well with the ground truth.
The latter shows that the phase errors were significantly increased by decreasing the observation points.
Here, the phase error was computed as follows:

\begin{eqnarray}
    \mathrm{err}(k) &= \left\langle|\mathrm{err}(\bm k')|\right\rangle_{k - 1/2 \leq |\bm k'| \leq k + 1/2}, \\
    \mathrm{err}(\bm k) &= \min |\arg \bm E^\mathrm{t}(\bm k) - \arg \bm E^\mathrm{DA}(\bm k) + 2n\pi| \ \forall n\in\{-1, 0, 1\},
\end{eqnarray}

\noindent
where $E^\mathrm{t}(\bm k)$ and $E^\mathrm{DA}(\bm k)$ are energy spectra of the ground truth and the result of the DA, respectively, and 
$n$ is introduced for periodicity.
The bracket $\langle val \rangle_{cond}$ computes the mean of the value $val$ over the range where the condition $cond$ is satisfied.

\figref{fig:spaghetti} is helpful to understand the uncertainty.
The figure shows the spaghetti plot, which visually demonstrates the uncertainty of the positions of eddies in the ensemble simulation.
The figure indicates the following insights: the results with $p=1,2,4,8,16, 32$ showed relatively small uncertainty, while those with $p=64$ showed the visibly large uncertainty in the small-scale vortices.
These findings are also consistent with the quantitative evaluation of the RMSE in \figref{fig:prune}. 

From these results, it was shown that in the sparse observation case, the disappearance of the small-scale vortices was not due to the numerical viscosity but was due to the ensemble average; this can be seen as a kind of Reynolds averaging over the ensemble simulation.
Overall, it was clarified that the LETKF is characterized by a weak numerical dissipation, and the amplitude of the energy spectrum is well maintained even with sparse observation, while the phase error or the time-dependent information is largely affected by the resolution of observation points.

Next, we discuss the cause of the numerical instability at small ensemble sizes,
which was shown in \figref{fig:prune}.
The origin of the instability might be the analysis increment, namely, 
the difference between the analysis and the forecast.
We evaluated the energy spectrum of the analysis increment,

\begin{equation}
    \Delta(k) = \frac{1}{M} \sum_{m=0}^{M-1}
        \sum_{||\bm k'|-k|\leq1/2} 
            \frac{1}{2}|\{\bm u_m^\mathrm{a} - \bm u_m^\mathrm{f}\}(\bm k')|^2.
\end{equation}

\noindent
\figref{fig:letkf_z_matrix} shows the results at the moderately sparse observation cases.
At $p=32$, only the case with $M=4$ was numerically unstable, and the corresponding $\Delta(k)$, which was observed just before the divergence, was larger than that with $M=16, 64$ in the whole wavenumber regime.
Especially, it was enhanced in the high wavenumber regime ($k\gtrapprox10$);
the amplitude of $\Delta(k)$ in the unstable case ($M=4$) was a few orders of magnitudes larger than that in the stable cases ($M=16, 64$).
In addition, the amplitude of $\Delta(k)$ was comparable to the energy spectrum $E(k)$ for $k\gtrapprox 30$.
A similar trend was also seen at $p=64$; the cases with $M=4, 16$ were numerically unstable, and they showed larger $\Delta(k)$ than the stable case ($M=64$) for $k\gtrapprox10$.
In the unstable cases, $\Delta(k)$ was comparable to $E(k)$ for $k\gtrapprox 30$, and was larger than $E(k)$ in the dissipation regime ($k\gtrapprox100$).
Hence, such large analysis increment in the high wavenumber regime was
the cause of the noisy gain, leading to the spurious energy injection in the high wavenumber regime, and thus the numerical instability.

\section{Conclusion}
\label{s:conclusion}

This paper investigated the ensemble data assimilation (DA) for the large eddy simulation (LES) by the lattice Boltzmann method coupled with the local ensemble transform Kalman filter (LBM-LETKF).
We showed that the LETKF enables a straightforward implementation in the LBM, where the velocity distribution function is assimilated based on the macroscopic observable variables such as the density and the velocity.
We compared two DA methods of the LETKF and the nudging by the observing system simulation experiment (OSSE) of the 2D forced isotropic turbulence simulation.
Firstly, in the spatially-dense observation case, both the nudging and the LETKF showed high DA accuracy, while the LETKF showed better DA accuracy than the nudging.
The duration of the transient period from non-assimilated to fully-assimilated state was also much faster in the LETKF than in the nudging.
Although the nudging resulted in a noisy field because of the observation error, the LETKF reproduced a quite accurate energy spectrum up to the high wavenumber regime, where the observation error was relatively large.
Secondly, in the spatially-sparse observation case, the nudging gave good results up to the observation points of $32\times32$ against the numerical resolution of $256\times256$.
The LETKF, in contrast, showed good results up to the observation points of $8\times8$, which is consistent with the Nyquist wavenumber needed to capture the energy-containing scale.
The analysis of the energy spectra also revealed the important features of the LETKF: 
\begin{itemize}
    \item The lack of observation points affected mainly the time-dependent information (i.e. the phase),
    as shown in \figref{fig:letkf_fft_argerr}. 
    The mean amplitude was well maintained, as seen in \figref{fig:prune_fft} (left). 
    \item The lack of observation points produced the spurious energy injection at the high wavenumber regime, leading to numerical instability.
    This phenomenon was known as catastrophic filter divergence~\citep{Harlim2010,Gottwald2013,Kelly2015}, which
    could be suppressed by increasing the ensemble size.
\end{itemize}
Overall, the LBM-LETKF is highly robust and accurate, provided that the ensemble size is large enough for the given configuration.
From these results, we conclude that the LBM-LETKF is an encouraging DA approach for the LES of turbulent flows at least in 2D problems.

In future work, we will study the applicability of the LBM-LETKF in 3D LESs.
For this purpose, we are going to address high-performance computing (HPC) techniques of the LBM-LETKF toward extreme-scale 3D LESs.

\section*{Acknowledgements}

Computations were performed on the NVIDIA DGX-2 system at Japan Atomic Energy Agency (JAEA), the HPE SGI8600 supercomputer at JAEA, and the Wisteria-Aquarius supercomputer at The University of Tokyo.

\section*{Funding information}
This work was funded by the JSPS KAKENHI [Grant Numbers JP21K17755, JP22H03599], and the Joint Usage/Research Center for Interdisciplinary Large-scale Information Infrastructures (JHPCN) [Project IDs jh210003, jh210049, jh220030, jh230014].

\section*{Declaration of Competing Interest}

The authors declare that they have no known competing financial interests or personal relationships that could have appeared to influence the work reported in this paper.

\section*{Data availability}

The software that provided results of this study is available on \\\url{https://github.com/hasegawa-yuta-jaea/LBM2D-LETKF}.
Output data are available upon reasonable requests.

\section*{Authors contribution}

\textbf{Yuta Hasegawa:} Methodology, Software, Validation, Formal analysis, Investigation, Data Curation, Writing - Original draft preparation.
\textbf{Naoyuki Onodera:} Software, Formal analysis.
\textbf{Yuuichi Asahi:} Formal analysis.
\textbf{Takuya Ina:} Software.
\textbf{Toshiyuki Imamura:} Software.
\textbf{Yasuhiro Idomura:} Conceptualization, Writing - Review \& Editing, Formal analysis, Supervision.

\newcommand{\newblock}{}
\bibliographystyle{jfm} 

\begin{thebibliography}{65}
\expandafter\ifx\csname natexlab\endcsname\relax\def\natexlab#1{#1}\fi
\def\au#1{#1} \def\ed#1{#1} \def\yr#1{#1}\def\at#1{#1}\def\jt#1{\textit{#1}}
  \def\bt#1{#1}\def\bvol#1{\textbf{#1}} \def\vol#1{#1} \def\pg#1{#1}
  \def\publ#1{#1}\def\arxiv#1{#1}\def\org#1{#1}\def\st#1{\textit{#1}}

\bibitem[Anderson(2007)]{Anderson2007a}
{\sc \au{Anderson, Jeffrey~L.}} \yr{2007}  \at{{An adaptive covariance
  inflation error correction algorithm for ensemble filters}}.  \jt{Tellus A:
  Dynamic Meteorology and Oceanography}  \bvol{59}~(2),  \pg{210--224}.

\bibitem[Anderson \& Anderson(1999)]{Anderson1999a}
{\sc \au{Anderson, Jeffrey~L.} \& \au{Anderson, Stephen~L.}} \yr{1999}  \at{{A
  Monte Carlo implementation of the nonlinear filtering problem to produce
  ensemble assimilations and forecasts}}.  \jt{Monthly Weather Review}
  \bvol{127}~(12),  \pg{2741--2758}.

\bibitem[Azouani {\em et~al.\/}(2014)Azouani, Olson \& Titi]{Azouani2014}
{\sc \au{Azouani, Abderrahim}, \au{Olson, Eric} \& \au{Titi, Edriss~S.}}
  \yr{2014}  \at{{Continuous Data Assimilation Using General Interpolant
  Observables}}.  \jt{Journal of Nonlinear Science}  \bvol{24}~(2),
  \pg{277--304}.

\bibitem[Batchelor(1969)]{Batchelor1969-hyphen}
{\sc \au{Batchelor, G.~K.}} \yr{1969}  \at{{Computation of the Energy Spectrum
  in Homogeneous Two‐Dimensional Turbulence}}.  \jt{The Physics of Fluids}
  \bvol{12}~(12),  \pg{II\hyphen233--II\hyphen239}.

\bibitem[Bauweraerts \& Meyers(2021)]{Bauweraerts2021}
{\sc \au{Bauweraerts, Pieter} \& \au{Meyers, Johan}} \yr{2021}
  \at{{Reconstruction of turbulent flow fields from lidar measurements using
  large-eddy simulation}}.  \jt{Journal of Fluid Mechanics}  \bvol{906},
  \pg{A17}.

\bibitem[Boffetta(2007)]{Boffetta2007}
{\sc \au{Boffetta, G.}} \yr{2007}  \at{{Energy and enstrophy fluxes in the
  double cascade of two-dimensional turbulence}}.  \jt{Journal of Fluid
  Mechanics}  \bvol{589},  \pg{253--260}.

\bibitem[Burgers {\em et~al.\/}(1998)Burgers, Van~Leeuwen \&
  Evensen]{Burgers1998}
{\sc \au{Burgers, Gerrit}, \au{Van~Leeuwen, Peter~Jan} \& \au{Evensen, Geir}}
  \yr{1998}  \at{{Analysis scheme in the ensemble Kalman filter}}.  \jt{Monthly
  Weather Review}  \bvol{126}~(6),  \pg{1719--1724}.

\bibitem[Chertkov {\em et~al.\/}(2007)Chertkov, Connaughton, Kolokolov \&
  Lebedev]{Chertkov2007}
{\sc \au{Chertkov, M.}, \au{Connaughton, C.}, \au{Kolokolov, I.} \&
  \au{Lebedev, V.}} \yr{2007}  \at{{Dynamics of energy condensation in
  two-dimensional turbulence}}.  \jt{Physical Review Letters}  \bvol{99}~(8),
  \pg{084501}.

\bibitem[Clark {\em et~al.\/}(2020)Clark, Tarra \& Berera]{Clark2020}
{\sc \au{Clark, Daniel}, \au{Tarra, Lukas} \& \au{Berera, Arjun}} \yr{2020}
  \at{{Chaos and information in two-dimensional turbulence}}.  \jt{Physical
  Review Fluids}  \bvol{5}~(6),  \pg{064608}.

\bibitem[Colburn {\em et~al.\/}(2011)Colburn, Cessna \& Bewley]{Colburn2011}
{\sc \au{Colburn, C.~H.}, \au{Cessna, J.~B.} \& \au{Bewley, T.~R.}} \yr{2011}
  \at{{State estimation in wall-bounded flow systems. Part 3. the ensemble
  Kalman filter}}.  \jt{Journal of Fluid Mechanics}  \bvol{682},
  \pg{289--303}.

\bibitem[Desroziers {\em et~al.\/}(2005)Desroziers, Berre, Chapnik \&
  Poli]{Desroziers2006}
{\sc \au{Desroziers, G.}, \au{Berre, L.}, \au{Chapnik, B.} \& \au{Poli, P.}}
  \yr{2005}  \at{{Diagnosis of observation, background and analysis-error
  statistics in observation space}}.  \jt{Quarterly Journal of the Royal
  Meteorological Society}  \bvol{131}~(613),  \pg{3385--3396}.

\bibitem[Evensen(1994)]{Evensen1994a}
{\sc \au{Evensen, Geir}} \yr{1994}  \at{{Sequential data assimilation with a
  nonlinear quasi-geostrophic model using Monte Carlo methods to forecast error
  statistics}}.  \jt{Journal of Geophysical Research}  \bvol{99}~(C5),
  \pg{143--162}.

\bibitem[Evensen(2003)]{Evensen2003a}
{\sc \au{Evensen, Geir}} \yr{2003}  \at{{The Ensemble Kalman Filter:
  Theoretical formulation and practical implementation}}.  \jt{Ocean Dynamics}
  \bvol{53}~(4),  \pg{343--367}.

\bibitem[Geier {\em et~al.\/}(2015)Geier, Sch{\"{o}}nherr, Pasquali \&
  Krafczyk]{Geier2015a}
{\sc \au{Geier, Martin}, \au{Sch{\"{o}}nherr, Martin}, \au{Pasquali, Andrea} \&
  \au{Krafczyk, Manfred}} \yr{2015}  \at{{The cumulant lattice Boltzmann
  equation in three dimensions: Theory and validation}}.  \jt{Computers {\&}
  Mathematics with Applications}  \bvol{70}~(4),  \pg{507--547}.

\bibitem[Gesho {\em et~al.\/}(2016)Gesho, Olson \& Titi]{Gesho2016}
{\sc \au{Gesho, Masakazu}, \au{Olson, Eric} \& \au{Titi, Edriss~S.}} \yr{2016}
  \at{{A Computational Study of a Data Assimilation Algorithm for the
  Two-dimensional Navier^^e2^^80^^93Stokes Equations}}.  \jt{Journal of Fluid
  Mechanics}  \bvol{19}~(4),  \pg{1094--1110}.

\bibitem[Gottwald \& Majda(2013)]{Gottwald2013}
{\sc \au{Gottwald, G.~A.} \& \au{Majda, A.~J.}} \yr{2013}  \at{{A mechanism for
  catastrophic filter divergence in data assimilation for sparse observation
  networks}}.  \jt{Nonlinear Processes in Geophysics}  \bvol{20}~(5),
  \pg{705--712}.

\bibitem[Harlim \& Majda(2010)]{Harlim2010}
{\sc \au{Harlim, John} \& \au{Majda, Andrew~J.}} \yr{2010}  \at{{Filtering
  turbulent sparsely observed geophysical flows}}.  \jt{Monthly Weather Review}
   \bvol{138}~(4),  \pg{1050--1083}.

\bibitem[Hasegawa {\em et~al.\/}(2022)Hasegawa, Imamura, Ina, Onodera, Asahi \&
  Idomura]{Hasegawa2022}
{\sc \au{Hasegawa, Yuta}, \au{Imamura, Toshiyuki}, \au{Ina, Takuya},
  \au{Onodera, Naoyuki}, \au{Asahi, Yuuichi} \& \au{Idomura, Yasuhiro}}
  \yr{2022} {GPU Optimization of Lattice Boltzmann Method with Local Ensemble
  Transform Kalman Filter}.  \bt{In {\em 2022 IEEE/ACM Workshop on Latest
  Advances in Scalable Algorithms for Large-Scale Heterogeneous Systems
  (ScalAH)\/}},  \pg{pp. 10--17}.

\bibitem[Herring \& Mcwilliams(1985)]{Herring1985}
{\sc \au{Herring, J.~R.} \& \au{Mcwilliams, J.~C.}} \yr{1985}  \at{{Comparison
  of direct numerical simulation of two-dimensional turbulence with two-point
  closure: The effects of intermittency}}.  \jt{Journal of Fluid Mechanics}
  \bvol{153},  \pg{229--242}.

\bibitem[Hoke \& Anthes(1976)]{Hoke1976}
{\sc \au{Hoke, James~E.} \& \au{Anthes, Richard~A.}} \yr{1976}  \at{{The
  Initialization of Numerical Models by a Dynamic-Initialization Technique}}.
  \jt{Monthly Weather Review}  \bvol{104},  \pg{1551--1556}.

\bibitem[Hunt {\em et~al.\/}(2007)Hunt, Kostelich \& Szunyogh]{Hunt2007a}
{\sc \au{Hunt, Brian~R.}, \au{Kostelich, Eric~J.} \& \au{Szunyogh, Istvan}}
  \yr{2007}  \at{{Efficient data assimilation for spatiotemporal chaos: A local
  ensemble transform Kalman filter}}.  \jt{Physica D: Nonlinear Phenomena}
  \bvol{230},  \pg{112--126}.

\bibitem[Kalnay(2002)]{Kalnay2002}
{\sc \au{Kalnay, Eugenia}} \yr{2002} {\em {Atmospheric Modeling, Data
  Assimilation and Predictability}\/}.  \publ{Cambridge University Press}.

\bibitem[Kelly {\em et~al.\/}(2015)Kelly, Majda \& Tong]{Kelly2015}
{\sc \au{Kelly, David}, \au{Majda, Andrew~J.} \& \au{Tong, Xin~T.}} \yr{2015}
  \at{{Concrete ensemble Kalman filters with rigorous catastrophic filter
  divergence}}.  \jt{Proceedings of the National Academy of Sciences of the
  United States of America}  \bvol{112}~(34),  \pg{10589--10594}.

\bibitem[Kraichnan(1967)]{Kraichnan1967}
{\sc \au{Kraichnan, Robert~H.}} \yr{1967}  \at{{Inertial ranges in
  two-dimensional turbulence}}.  \jt{The Physics of Fluids}  \bvol{10}~(7),
  \pg{1417--1423}.

\bibitem[Kr{\"{u}}ger {\em et~al.\/}(2017)Kr{\"{u}}ger, Kusumaatmaja, Kuzmin,
  Shardt, Silva \& Viggen]{Kruger2017}
{\sc \au{Kr{\"{u}}ger, Timm}, \au{Kusumaatmaja, Halim}, \au{Kuzmin, Alexandr},
  \au{Shardt, Orest}, \au{Silva, Goncalo} \& \au{Viggen, Erlend~Magnus}}
  \yr{2017} {\em {The Lattice Boltzmann Method: Principles and Practice}\/}.
  \publ{Springer}.

\bibitem[Labahn {\em et~al.\/}(2020)Labahn, Wu, Harris, Coriton, Frank \&
  Ihme]{Labahn2020}
{\sc \au{Labahn, Jeffrey~W.}, \au{Wu, Hao}, \au{Harris, Shaun~R.}, \au{Coriton,
  Bruno}, \au{Frank, Jonathan~H.} \& \au{Ihme, Matthias}} \yr{2020}
  \at{{Ensemble Kalman Filter for Assimilating Experimental Data into
  Large-Eddy Simulations of Turbulent Flows}}.  \jt{Flow, Turbulence and
  Combustion}  \bvol{104}~(4),  \pg{861--893}.

\bibitem[Lakshmivarahan \& Lewis(2013)]{Lakshmivarahan2013-hyphen}
{\sc \au{Lakshmivarahan, S} \& \au{Lewis, John~M}} \yr{2013}  \at{{Nudging
  Methods: A Critical Overview}}.  \bt{In {\em Data Assimilation for
  Atmospheric, Oceanic and Hydrologic Applications (Vol. II)\/}},  \pg{pp.
  27--57}.  \publ{Berlin, Heidelberg: Springer}.

\bibitem[Lalescu {\em et~al.\/}(2013)Lalescu, Meneveau \& Eyink]{Lalescu2013}
{\sc \au{Lalescu, Cristian~C.}, \au{Meneveau, Charles} \& \au{Eyink,
  Gregory~L.}} \yr{2013}  \at{{Synchronization of Chaos in Fully Developed
  Turbulence}}.  \jt{Physical Review Letters}  \bvol{110},  \pg{084102}.

\bibitem[Le~Dimet \& Talagrand(1986)]{LeDimet1986}
{\sc \au{Le~Dimet, F.~X.} \& \au{Talagrand, O.}} \yr{1986}  \at{{Variational
  algorithms for analysis and assimilation of meteorological observations:
  theoretical aspects.}}  \jt{Tellus A: Dynamic Meteorology and Oceanography}
  \bvol{38}~(2),  \pg{97--110}.

\bibitem[Legras {\em et~al.\/}(1988)Legras, Santangelo \& Benzi]{Legras1988}
{\sc \au{Legras, B}, \au{Santangelo, P} \& \au{Benzi, R}} \yr{1988}
  \at{{High-Resolution Numerical Experiments for Forced Two-Dimensional
  Turbulence}}.  \jt{Europhysics Letters (EPL)}  \bvol{5}~(1),  \pg{37--42}.

\bibitem[Leith(1968)]{Leith1968}
{\sc \au{Leith, C.E.}} \yr{1968}  \at{{Diffusion Approximation for
  Two-dimensional Turbulence}}.  \jt{The Physics of Fluids}  \bvol{11},
  \pg{671--673}.

\bibitem[Leoni {\em et~al.\/}(2020)Leoni, Mazzino \& Biferale]{Leoni2020}
{\sc \au{Leoni, Patricio Clark~Di}, \au{Mazzino, Andrea} \& \au{Biferale,
  Luca}} \yr{2020}  \at{{Synchronization to Big Data: Nudging the Navier-Stokes
  Equations for Data Assimilation of Turbulent Flows}}.  \jt{Physical Review X}
   \bvol{10}~(1),  \pg{011023}.

\bibitem[Li {\em et~al.\/}(2022)Li, Tian \& Li]{Li2022a}
{\sc \au{Li, Jian}, \au{Tian, Mengdan} \& \au{Li, Yi}} \yr{2022}
  \at{{Synchronizing large eddy simulations with direct numerical simulations
  via data assimilation}}.  \jt{Physics of Fluids}  \bvol{34}~(6),
  \pg{065108}.

\bibitem[Li {\em et~al.\/}(2012)Li, Luo, He, Gao \& Tao]{Li2012}
{\sc \au{Li, Q.}, \au{Luo, K.~H.}, \au{He, Y.~L.}, \au{Gao, Y.~J.} \& \au{Tao,
  W.~Q.}} \yr{2012}  \at{{Coupling lattice Boltzmann model for simulation of
  thermal flows on standard lattices}}.  \jt{Physical Review E}  \bvol{85}~(1),
   \pg{1--16}.

\bibitem[Lilly(1969)]{Lilly1969}
{\sc \au{Lilly, Douglas~K}} \yr{1969}  \at{{Numerical Simulation of
  Two-Dimensional Turbulence}}.  \jt{The Physics of Fluids}  \bvol{12},
  \pg{II{\_}240--II{\_}249}.

\bibitem[Miyoshi(2011)]{Miyoshi2011}
{\sc \au{Miyoshi, Takemasa}} \yr{2011}  \at{{The gaussian approach to adaptive
  covariance inflation and its implementation with the local ensemble transform
  Kalman filter}}.  \jt{Monthly Weather Review}  \bvol{139}~(5),
  \pg{1519--1535}.

\bibitem[Miyoshi {\em et~al.\/}(2014)Miyoshi, Kondo \& Imamura]{Miyoshi2014}
{\sc \au{Miyoshi, Takemasa}, \au{Kondo, Keiichi} \& \au{Imamura, Toshiyuki}}
  \yr{2014}  \at{{The 10,240-member ensemble Kalman filtering with an
  intermediate AGCM}}.  \jt{Geophysical Research Letters}  \bvol{41}~(14),
  \pg{5264--5271}.

\bibitem[Miyoshi {\em et~al.\/}(2016)Miyoshi, Kunii, Ruiz, Lien, Satoh, Ushio,
  Bessho, Seko, Tomita \& Ishikawa]{Miyoshi2016a}
{\sc \au{Miyoshi, Takemasa}, \au{Kunii, Masaru}, \au{Ruiz, Juan}, \au{Lien,
  Guo~Yuan}, \au{Satoh, Shinsuke}, \au{Ushio, Tomoo}, \au{Bessho, Kotaro},
  \au{Seko, Hiromu}, \au{Tomita, Hirofumi} \& \au{Ishikawa, Yutaka}} \yr{2016}
  \at{{``Big data assimilation'' revolutionizing severe weather prediction}}.
  \jt{Bulletin of the American Meteorological Society}  \bvol{97}~(8),
  \pg{1347--1354}.

\bibitem[Miyoshi {\em et~al.\/}(2007)Miyoshi, Yamane \& Enomoto]{Miyoshi2007}
{\sc \au{Miyoshi, Takemasa}, \au{Yamane, Shozo} \& \au{Enomoto, Takeshi}}
  \yr{2007}  \at{{Localizing the Error Covariance by Physical Distances within
  a Local Ensemble Transform Kalman Filter (LETKF)}}.  \jt{Sola}  \bvol{3}~(1),
   \pg{89--92}.

\bibitem[Molteni(2003)]{Molteni2003}
{\sc \au{Molteni, F.}} \yr{2003}  \at{{Atmospheric simulations using a GCM with
  simplified physical parametrizations. I: Model climatology and variability in
  multi-decadal experiments}}.  \jt{Climate Dynamics}  \bvol{20}~(2-3),
  \pg{175--191}.

\bibitem[Nastac {\em et~al.\/}(2017)Nastac, Labahn, Magri \& Ihme]{Nastac2017}
{\sc \au{Nastac, Gabriel}, \au{Labahn, Jeffrey~W.}, \au{Magri, Luca} \&
  \au{Ihme, Matthias}} \yr{2017}  \at{{Lyapunov exponent as a metric for
  assessing the dynamic content and predictability of large-eddy simulations}}.
   \jt{Physical Review Fluids}  \bvol{2}~(9),  \pg{094606}.

\bibitem[Olson \& Titi(2003)]{Olson2003}
{\sc \au{Olson, Eric} \& \au{Titi, Edriss~S}} \yr{2003}  \at{{Determining Modes
  for Continuous Data Assimilation in 2D Tuebulence}}.  \jt{Journal of
  Statistical Physics}  \bvol{113}~(5),  \pg{799--840}.

\bibitem[Olson \& Titi(2008)]{Olson2008}
{\sc \au{Olson, Eric} \& \au{Titi, Edriss~S.}} \yr{2008}  \at{{Determining
  modes and Grashof number in 2D turbulence: A numerical case study}}.
  \jt{Theoretical and Computational Fluid Dynamics}  \bvol{22}~(5),
  \pg{327--339}.

\bibitem[Onodera {\em et~al.\/}(2018)Onodera, Idomura, Ali \&
  Shimokawabe]{Onodera2018}
{\sc \au{Onodera, Naoyuki}, \au{Idomura, Yasuhiro}, \au{Ali, Yussuf} \&
  \au{Shimokawabe, Takashi}} \yr{2018} {Communication Reduced Multi-time-step
  Algorithm for Real-time Wind Simulation on GPU-based Supercomputers}.  \bt{In
  {\em ScalA 2018: Proceedings of 2018 IEEE/ACM 9th Workshop on Latest Advances
  in Scalable Algorithms for Large-Scale Systems (ScalA)\/}},  \pg{pp. 9--16}.

\bibitem[Onodera {\em et~al.\/}(2021)Onodera, Idomura, Hasegawa, Nakayama,
  Shimokawabe \& Aoki]{Onodera2021}
{\sc \au{Onodera, Naoyuki}, \au{Idomura, Yasuhiro}, \au{Hasegawa, Yuta},
  \au{Nakayama, Hiromasa}, \au{Shimokawabe, Takashi} \& \au{Aoki, Takayuki}}
  \yr{2021}  \at{{Real-time tracer dispersion simulation in Oklahoma City using
  locally-mesh refined lattice Boltzmann method}}.  \jt{Boundary-Layer
  Meteorology}  \bvol{179},  \pg{187--208}.

\bibitem[Qian {\em et~al.\/}(1992)Qian, D’Humi{\`{e}}res \&
  Lallemand]{Qian1992}
{\sc \au{Qian, Y.~H.}, \au{D’Humi{\`{e}}res, D.} \& \au{Lallemand, P.}}
  \yr{1992}  \at{{Lattice BGK models for navier-stokes equation}}.
  \jt{Europhysics Letters}  \bvol{17}~(6),  \pg{479--484}.

\bibitem[Roussel {\em et~al.\/}(2013)Roussel, Bourgois, Benjelloun \&
  Delmaire]{Roussel2013}
{\sc \au{Roussel, Gilles}, \au{Bourgois, Laurent}, \au{Benjelloun, Mohammed} \&
  \au{Delmaire, Gilles}} \yr{2013}  \at{{Estimation of a semi-physical GLBE
  model using dual EnKF learning algorithm coupled with a sensor network design
  strategy: application to air field monitoring}}.  \jt{Information Fusion}
  \bvol{14}~(4),  \pg{335--348}.

\bibitem[Salman {\em et~al.\/}(2022)Salman, Khan, Kemp \& Noakes]{Salman2022}
{\sc \au{Salman, N.}, \au{Khan, A.}, \au{Kemp, A.~H.} \& \au{Noakes, C.~J.}}
  \yr{2022}  \at{{Indoor Temperature Forecast based on the Lattice Boltzmann
  method and Data Assimilation}}.  \jt{Building and Environment}  \bvol{210},
  \pg{108654}.

\bibitem[Scott(2007)]{Scott2007}
{\sc \au{Scott, R.~K.}} \yr{2007}  \at{{Nonrobustness of the two-dimensional
  turbulent inverse cascade}}.  \jt{Physical Review E}  \bvol{75}~(4),
  \pg{046301}.

\bibitem[Smagorinsky(1963)]{Smagorinsky1963a}
{\sc \au{Smagorinsky, J.}} \yr{1963}  \at{{General Circulation Experiments with
  the Primitive Equations: I. The Basic Experiment}}.  \jt{Monthly Weather
  Review}  \bvol{91}~(3),  \pg{99--164}.

\bibitem[Smith \& Yakhot(1994)]{Smith1994}
{\sc \au{Smith, Leslie~M.} \& \au{Yakhot, Victor}} \yr{1994}  \at{{Finite-Size
  Effects in Forced Two-Dimensional Turbulence}}.  \jt{Journal of Fluid
  Mechanics}  \bvol{274},  \pg{115--138}.

\bibitem[Suzuki(2012)]{Suzuki2012}
{\sc \au{Suzuki, Takao}} \yr{2012}  \at{{Reduced-order Kalman-filtered hybrid
  simulation combining particle tracking velocimetry and direct numerical
  simulation}}.  \jt{Journal of Fluid Mechanics}  \bvol{709},  \pg{249--288}.

\bibitem[Suzuki \& Hasegawa(2017)]{Suzuki2017}
{\sc \au{Suzuki, Takao} \& \au{Hasegawa, Yosuke}} \yr{2017}  \at{{Estimation of
  turbulent channel flow at $Re_\tau$=100 based on the
  wall measurement using a simple sequential approach}}.  \jt{Journal of Fluid
  Mechanics}  \bvol{830},  \pg{760--796}.

\bibitem[Szunyogh {\em et~al.\/}(2005)Szunyogh, Kostelich, Gyarmati, Patil,
  Hunt, Kalnay, Ott \& Yorke]{Szunyogh2005a}
{\sc \au{Szunyogh, Istvan}, \au{Kostelich, Eric~J.}, \au{Gyarmati, G.},
  \au{Patil, D.~J.}, \au{Hunt, Brian~R.}, \au{Kalnay, Eugenia}, \au{Ott,
  Edward} \& \au{Yorke, James~A.}} \yr{2005}  \at{{Assessing a local ensemble
  Kalman filter: perfect model experiments with the National Centers for
  Environmental Prediction global model}}.  \jt{Tellus A: Dynamic Meteorology
  and Oceanography}  \bvol{57}~(4),  \pg{528}.

\bibitem[Talagrand(1997)]{Talagrand1997}
{\sc \au{Talagrand, Oliver}} \yr{1997}  \at{{Assimialtion of Observation, an
  Introduction}}.  \jt{Journal of the Meteorological Societry of Japan}
  \bvol{75}~(1B),  \pg{191--209}.

\bibitem[Tippett {\em et~al.\/}(2003)Tippett, Anderson, Bishop, Hamill \&
  Whitaker]{Tippett2003}
{\sc \au{Tippett, Michael~K.}, \au{Anderson, Jeffrey~L.}, \au{Bishop,
  Craig~H.}, \au{Hamill, Thomas~M.} \& \au{Whitaker, Jeffrey~S.}} \yr{2003}
  \at{{Ensemble square root filters}}.  \jt{Monthly Weather Review}
  \bvol{131}~(7),  \pg{1485--1490}.

\bibitem[Tran \& Bowman(2003)]{Tran2003}
{\sc \au{Tran, Chuong~V.} \& \au{Bowman, John~C.}} \yr{2003}  \at{{On the dual
  cascade in two-dimensional turbulence}}.  \jt{Physica D: Nonlinear Phenomena}
   \bvol{176}~(3-4),  \pg{242--255}.

\bibitem[Tran \& Shepherd(2002)]{Tran2002}
{\sc \au{Tran, Chuong~V.} \& \au{Shepherd, Theodore~G.}} \yr{2002}
  \at{{Constraints on the spectral distribution of energy and enstrophy
  dissipation in forced two-dimensional turbulence}}.  \jt{Physica D: Nonlinear
  Phenomena}  \bvol{165}~(3-4),  \pg{199--212}.

\bibitem[Tsang \& Young(2009)]{Tsang2009}
{\sc \au{Tsang, Yue~Kin} \& \au{Young, William~R.}} \yr{2009}
  \at{{Forced-dissipative two-dimensional turbulence: A scaling regime
  controlled by drag}}.  \jt{Physical Review E}  \bvol{79}~(4),
  \pg{045308(R)}.

\bibitem[Wang \& Zaki(2022)]{Wang2022a}
{\sc \au{Wang, Mengze} \& \au{Zaki, Tamer~A.}} \yr{2022}  \at{{Synchronization
  of turbulence in channel flow}}.  \jt{Journal of Fluid Mechanics}
  \bvol{943},  \pg{A4}.

\bibitem[Wang {\em et~al.\/}(2022)Wang, Yuan, Xie \& Wang]{Wang2022}
{\sc \au{Wang, Yunpeng}, \au{Yuan, Zelong}, \au{Xie, Chenyue} \& \au{Wang,
  Jianchun}} \yr{2022}  \at{{Temporally sparse data assimilation for the
  small-scale reconstruction of turbulence}}.  \jt{Physics of Fluids}
  \bvol{34}~(6),  \pg{065115}.

\bibitem[Xie(2020)]{Xie2020}
{\sc \au{Xie, Jin~Han}} \yr{2020}  \at{{Quantifying the linear damping in
  two-dimensional turbulence}}.  \jt{Physical Review Fluids}  \bvol{5}~(9),
  \pg{094605}.

\bibitem[Yashiro {\em et~al.\/}(2020)Yashiro, Terasaki, Kawai, Kudo, Miryoshi,
  Imamura, Minami, Inoue, Tatsuo, Saji, Sayoh \& Tomita]{Yashiro2020}
{\sc \au{Yashiro, Hisashi}, \au{Terasaki, Koji}, \au{Kawai, Yuta}, \au{Kudo,
  Shuhei}, \au{Miryoshi, Takemasa}, \au{Imamura, Toshiyuki}, \au{Minami,
  Kazuo}, \au{Inoue, Hikaru}, \au{Tatsuo, Nishiki}, \au{Saji, Takayuki},
  \au{Sayoh, Masaki} \& \au{Tomita, Hirofumi}} \yr{2020} {A 1024-Member
  Ensemble Data Assimilation with 3. 5-Km Mesh Global Weather Simulations}.
  \bt{In {\em SC20: International Conference for High Performance Computing,
  Networking, Storage and Analysis\/}},  \pg{pp. 1--10}.

\bibitem[Yoshida {\em et~al.\/}(2005)Yoshida, Yamaguchi \& Kaneda]{Yoshida2005}
{\sc \au{Yoshida, Kyo}, \au{Yamaguchi, Junzo} \& \au{Kaneda, Yukio}} \yr{2005}
  \at{{Regeneration of Small Eddies by Data Assimilation in Turbulence}}.
  \jt{Physical Review Letters}  \bvol{94}~(1),  \pg{014501}.

\bibitem[Zauner {\em et~al.\/}(2022)Zauner, Mons, Marquet \&
  Leclaire]{Zauner2022}
{\sc \au{Zauner, M.}, \au{Mons, V.}, \au{Marquet, O.} \& \au{Leclaire, B.}}
  \yr{2022}  \at{{Nudging-based data assimilation of the turbulent flow around
  a square cylinder}}.  \jt{Journal of Fluid Mechanics}  \bvol{937},  \pg{A38}.

\end{thebibliography}
\providecommand*\hyphen{-}

\end{document}